\documentclass[fleqn,usenatbib]{mnras}
\usepackage[utf8]{inputenc}
\usepackage{newtxtext,newtxmath}
\DeclareUnicodeCharacter{2212}{-} 

\usepackage[T1]{fontenc}
\usepackage{comment}
\usepackage[normalem]{ulem}
\usepackage{xcolor}

\DeclareRobustCommand{\VAN}[3]{#2}
\let\VANthebibliography\thebibliography
\def\thebibliography{\DeclareRobustCommand{\VAN}[3]{##3}\VANthebibliography}

\usepackage{graphicx}	
\usepackage{amsmath}	
\usepackage{textcomp}   
\usepackage{soul}
\usepackage{siunitx}    
\usepackage{orcidlink}
\usepackage{overpic} 
\usepackage{tikz}
\usetikzlibrary{positioning}
\usepackage{fix-cm}

\title[Pre-flare Diagnostics by Aditya-L1]{Multi-Wavelength Diagnostics of Pre-Flare Evolution with Aditya-L1: From the Solar Chromosphere to the Corona}

\author[Adithya H. N.]{
Adithya H.N.,$^{1}$\orcidlink{0009-0002-1177-9948}\thanks{E-mail: adithya.mcnsmpl2023@learner.manipal.edu},
Sreejith Padinhatteeri$^{1}$\orcidlink{0000-0002-7276-4670}\thanks{E-mail: sreejith.p@manipal.edu},
Soumya Roy $^{1,2}$\orcidlink{0000-0003-2215-7810},
 K. Sankarasubramanian  $^{3,4}$\orcidlink{0000-0003-1406-4200},
 Durgesh Tripathi $^{5}$\orcidlink{0000-0003-1689-6254},
\newauthor Abhilash R Sarwade $^{3}$\orcidlink{0009-0002-0216-0545}, 
 Srikar Paavan Tadepalli $^{3}$\orcidlink{0009-0006-6197-9672},
 Harshavardhan G Hegde $^{6}$\orcidlink{0009-0005-4821-1039},
 Nived V. N $^{5}$\orcidlink{0000-0001-6866-6608},
 \newauthor Janmejoy Sarkar $^{5,7}$\orcidlink{0000-0002-8560-318X},
 Rahul Gopalakrishnan $^{5}$\orcidlink{0000-0002-1282-3480},
 Rushikesh Deogaonkar$^{5}$\orcidlink{},
A. N. Ramaprakash $^{5}$\orcidlink{},
 \newauthor Sami K. Solanki $^{7}$\orcidlink{0000-0002-3418-8449},
 Dibyendu Nandy $^{4,8}$\orcidlink{0000-0001-5205-2302},
Dipankar Banerjee $^{4,9}$\orcidlink{0000-0003-4653-6823}
\\
$^{1}$ Manipal Centre for Natural Sciences, Manipal Academy of Higher Education, Manipal, Karnataka, 576104, India\\
$^{2}$ Physical Research Laboratory, Navrangpura, Ahmedabad, Gujarat-380009, India \\
$^{3}$ U R Rao Satellite Centre, Old Airport Road, Vimanapura Post, Bengaluru - 560017, Karnataka, India\\
$^{4}$ Centre for Excellence in Space Sciences India, Indian Institute of Science Education and Research Kolkata, Mohanpur 741246, West Bengal, India \\
$^{5}$ Inter-University Centre for Astronomy and Astrophysics, Post Bag 4, Ganeshkhind, Pune- 411007, Maharashtra, India\\
$^{6}$ Sri Dharmasthala Manjunatheshwara College (Autonomous), Ujire-574240, India \\
$^{7}$ Max-Planck-Institut für Sonnensystemforschung, Justus-von-Liebig-Weg 3, 37077 Göttingen, Germany \\
$^{8}$ Department of Physical Sciences, Indian Institute of Science Education and Research Kolkata, Mohanpur 741246, West Bengal, India \\
$^{9}$ Indian Institute of Space Science and Technology, Valiamala, Thiruvananthapuram 695 547, Kerala, India 
}

\date{Accepted XXX. Received YYY; in original form ZZZ}

\pubyear{\the\year{}}

\begin{document}
\label{firstpage}
\pagerange{\pageref{firstpage}--\pageref{lastpage}}
\maketitle
\begin{abstract}
The pre-flare phase of solar flares provides important insight into the processes that drive active regions toward instability. We investigate chromospheric pre-flare activity using observations from the Solar Ultraviolet Imaging Telescope (SUIT) onboard Aditya-L1, complemented with X-ray measurements from  High Energy L1 Orbiting X-ray
Spectrometer (HEL1OS) and Solar Low Energy X-ray Spectrometer (SoLEXS). We analyse seven M- and X-class flares, focusing on spatially resolved \ion{Mg}{ii} h (2803~\AA) observations from SUIT.
We identify 102 pre-flare transients within regions of interest prior to flare onset. These transients are detected in the \ion{Mg}{ii} h channel, with no counterparts in continuum filters, confirming their chromospheric origin. In most cases, the transients are co-spatial with polarity inversion lines (PILs) and the eventual flaring region. Approximately 28~\% of transients have X-ray counterparts in HEL1OS (10-30~keV);  The Spectrometer Telescope for Imaging X-rays (STIX) spectral analysis reveals non-thermal emission in a subset, indicating that some transients are small-scale flare-like events.

A hot X-ray onset is identified in 
four cases. For the remaining three cases, the signal-to-noise ratio above the background is insufficient to determine whether a hot-onset phase is present.
The peak-flux distribution of the transients follows a broken power law with indices $\alpha_1 = 1.64^{+0.59}_{-0.57}$ and $\alpha_2 = 3.12^{+0.64}_{-0.61}$, with the higher-energy slope consistent with the Ly-$\alpha$ flare distribution. These results suggest that chromospheric pre-flare transients represent small-scale magnetic energy-release events that contribute to the progressive destabilisation of active regions prior to major flare onset.
\end{abstract}


\section{Introduction}
Solar flares are localised bursts of energy release that produce emission across the electromagnetic spectrum from multiple layers of the solar atmosphere. The chromosphere accounts for the majority of the radiated
energy during solar flares, and hence studying the chromosphere is important for understanding flare energetics and the precursors  \citep[see e.g.,][]{fletcher_chromosphere_2010, milligan_radiated_2014}. 
The exact physical process that initiates a flare is not yet fully understood; it may involve multiple processes rather than a single event. 
Indirect observations, such as pre-flare brightening, could serve as an indicator of flare triggers \citep{green_origin_2018}, here, 'trigger' refers to any localised process that initiates the flare.
Numerous studies of short-lived pre-flare brightenings in the chromosphere \citep{Hahn2005, Bamba2013, Bamba2018, dissauer_uniqueness_2025, kumar_compact_2026, Bamba2014, wangHighresolutionObservationsFlare2017b} 
and corona \citep{webb_coronal_1985, tappin_all_1991, farnik_spatial_1996, farnik_soft_1998, ChiMT_2006, Chifor2007, shohinUltravioletXrayPrecursors2024} have provided detailed insights into these phenomena and their association with solar flares.
Using \ion{Ca}{ii} H observations from the Solar Optical Telescope \citep[SOT;][]{Tsuneta2008} onboard Hinode, \citet{Kusano2012} and \citet{Bamba2013} observed that the small-scale transient brightenings occurring at PILs are the signatures of flare triggers.
\cite{dissauer_uniqueness_2025} analysed pre-flare activity using the Atmospheric Imaging Assembly \citep[AIA;][]{Lemen2012} 1600 {\AA} channel, revealing multiple transient brightenings clustered around strong PILs and regions of enhanced magnetic energy density. The AIA 1600 {\AA} channel samples emission from the upper photosphere to the transition region \citep{simoes_spectral_2019}.
\cite{woodsUnsupervisedMachineLearning2021} and  \cite{panosIdentifyingPreflareSpectral2023} reported intensity enhancements in the  \ion{Mg}{ii} triplet during the pre-flare phase. Studies using H$\alpha$ and \ion{Ca}{ii} K observations by \cite{wangHighresolutionObservationsFlare2017b} and \cite{kumar_compact_2026} suggest that these pre-flare brightenings are small-scale reconnections in the chromosphere. 
Despite these advances in pre-flare brightening studies, there are very few systematic multi-event studies that combine observations of the chromosphere and corona. Such studies are required to understand the energetics of these transients and their association with solar flares. 

How magnetised plasma systems accumulate energy through stress, and how this stored energy is released, has been an intriguing theoretical question with implications for astrophysical to laboratory plasmas. Hypotheses ranging from Taylor relaxation of twisted magnetic fields \citep{Taylor1974} to self-organised criticality and avalanches \citep{Bak1987}, e.g., the sand pile model, are relevant in this context, and have been applied to understand and interpret energy release in solar flares \citep{Lu1991, Nandy2003, Chitta2026}. Solar flare energies and their occurrence rates are known to follow power-law distributions, indicating scale-invariant behaviour of energy release in the solar atmosphere (e.g. \citealt{hudsonSolarFlaresMicroflares1991, aschwanden25YearsSelfOrganized2016, biasiottiStatisticalAnalysisSolar2025, liPowerlawDistributionsXray2016a}). Several studies, using Geostationary Operational Environmental Satellite (GOES) observations, report power-law indices in the range of $\mathrm{\alpha_{peak~irrad} \approx}$ 1.98 {–} 2.16 for peak X-ray flare energies \citep{masonCoronalHeatingDetermined2023, veronigTemporalAspectsFrequency2002a, aschwandenAutomatedSolarFlare2012}. Recent studies with the X-ray Solar Monitor \citep[XSM;][]{shanmugamSolarXrayMonitor2020} onboard Chandrayaan-2 also indicate the power law to be 1.96 \citep{Valluvan2024}). This behaviour is consistent with the framework of self-organised criticality (SOC) models of the solar corona \citep{aschwanden25YearsSelfOrganized2016, Lu1991}. 

Extending such analyses to lower atmosphere energetics, \cite{milligan_lyman-alpha_2020} investigated chromospheric flare distributions in Ly$\alpha$ using  EUV Sensor \citep[EUVS;][]{evansEarlyObservationsGOES132010,viereckSolarExtremeUltraviolet2007} on the GOES-15 satellite. They reported a steeper power-law index of 2.82 $\pm$ 0.27. These results suggest that small-scale energy-release events, such as microflares and nanoflares, may represent scaled-down versions of larger flares and arise from similar physical processes \citep{parkerNanoflaresSolarXRay1988, hannahRHESSIMicroflareStatistics2008a, hannahMicroflaresStatisticsXray2011,upendranImpulsiveHeatingQuiet2021,upendranNanoflareHeatingSolar2022,klimchukSolvingCoronalHeating2006}. Such small-scale events have also been proposed as potential contributors to coronal heating. 
In the pre-flare phase, such small-scale transient events are observed in the chromosphere.
However, the statistical behaviour of such chromospheric pre-flare transients remains poorly explored. 
Determining whether these pre-flare transients follow similar scaling laws is essential for understanding their role in the build-up and release of magnetic energy in active regions.

Recent studies using GOES/X-ray Sensor (XRS) observations have identified a hot X-ray onset phase, a few minutes prior to the impulsive phase of solar flares \citep{hudson_hot_2021, da_silva_statistical_2023, battaglia_existence_2023, hudson_anticipating_2025, hudson_hot_2025, telikicherla_improving_2025}. The hot-onset is generally characterised by an elevated temperature of $\approx$ 10–15 MK and a gradual increase in emission measure (EM). These emissions are predominantly thermal, suggesting that the physical processes differ from those of the typical flare impulsive phase \citep{hudson_hot_2021, battaglia_existence_2023}. Further insights from soft X-ray spectroscopy using the Dual-zone Aperture X-ray Solar Spectrometer \citep[DAXSS;][]{woodsNewSolarIrradiance2017} show that both temperature and elemental abundances evolve during the pre-flare phase  \citep{telikicherla_investigating_2024}.
In this context, Aditya-L1 \citep{seethaAdityaL1Mission2017, TriCNR_2023, Sankarasubramanian2025, Parate2025} provides a unique observational capability by combining near-ultraviolet (NUV) imaging with X-ray diagnostics, enabling a comprehensive investigation of the pre-flare phase. The Solar Ultraviolet Imaging Telescope \citep[SUIT;][]{tripathi_solar_2025, sarkar_test_2025} onboard Aditya-L1 observes the solar atmosphere from the upper photosphere to the chromosphere using 11 science filters, including three lines with chromospheric emission cores, namely, \ion{Mg}{ii} h, \ion{Mg}{ii} k, and \ion{Ca}{ii} H.

The \ion{Mg}{ii} h and k lines are highly sensitive to localised chromospheric heating and have been shown to respond strongly to small-scale energy-release events, making them particularly suitable for studying pre-flare activity \citep{litwickaStatisticalAnalysisCompact2025}. Simultaneous X-ray observations \citep{Sankarasubramanian2017} are provided by the Solar Low Energy X-ray Spectrometer \citep[SoLEXS;][]{sankarasubramanian_solar_2025} and the High Energy L1 Orbiting X-ray Spectrometer \citep[HEL1OS;][]{nandi_hel1os_2025}. SoLEXS, covering an energy range of 2 {--}22~keV, enables the study of thermal plasma in the corona by estimating temperature and emission measure, while HEL1OS probes the higher-energy, non-thermal regime using both CdTe and CZT detectors, with an energy range of 8{--}150~keV.

In this study, we investigate, for the first time, the pre-flare phases of multiple flares using spatially resolved NUV observations from SUIT, together with sun{--}as{--}a{--}star X-ray observations from SoLEXS and HEL1OS. 

We address the following key questions:

\begin{enumerate}
\item What is the behaviour of the NUV emission during the pre-flare phase?
\item Do chromospheric transients exhibit corresponding X-ray counterparts?
\item What is the nature of the pre-impulsive energy release during the flare onset phase, and how does the hot X-ray onset relate to the chromospheric activity observed in \ion{Mg}{ii} h?
\end{enumerate}

The rest of the paper is organised as follows. In Section~\ref{sec:observations}, we describe the observational data sets used in this study. Section~\ref{sec:methodology} outlines the methodology adopted to identify and characterise pre-flare transients and to analyse their multi-wavelength counterparts. In Section~\ref{sec:results}, we present the results for individual case studies, followed by a statistical analysis of the detected transients. Section~\ref{sec:discussion} discusses the spatial distribution, X-ray associations, and physical implications of the observed pre-flare activity. Finally, Section~\ref{sec:conclusions} summarises the main findings of this work.

\section{Observations}
\label{sec:observations}
In this study, we investigate the pre-flare phase of solar flares using multiple instruments onboard Aditya-L1, which together provide simultaneous observations of the chromosphere and the corona. SUIT observes the upper photosphere and chromosphere through 11 science filters, consisting of three broadband filters (BB01--BB03) and eight narrowband filters (NB01--NB08) \citep[see][for details on the SUIT instrument]{tripathi_solar_2025, sarkar_test_2025}. For the present analysis, we use six SUIT narrow-band filters. The NB01 filter (2140~\AA) is excluded due to significant throughput degradation caused by contamination within the telescope optical cavity \citep[see][]{gopalakrishnan_2025}. The NB08 filter (\ion{Ca}{II} H, 3968.5~\AA) data have not been included due to incomplete ghost and stray-light corrections \citep[see][]{sarkar_test_2025}. 
We analyse SUIT Level-1 Region of Interest (ROI) images ($\approx~$491\arcsec $\times$491\arcsec ), with a cadence of approximately 90~s during this observation period. 
These data are corrected for instrumental scatter and vignetting \cite[details of the calibration procedure are described in][]{sarkar_test_2025}.

To characterise the coronal emission, we use observations from SoLEXS (2–22 keV) and HEL1OS (CdTe; 8–70 keV), both moderate-resolution (SoLEXS; 170~ev at 5.9~keV , CdTe/HEL1OS; $\approx$ 1~keV at 14~keV) spectrographs that continuously observe the Sun as a star from the Sun–Earth L1 point. In addition, X-ray spectral analysis is performed using data from the Spectrometer Telescope for Imaging X-rays \citep[STIX;][]{krucker_spectrometertelescope_2020} onboard Solar Orbiter \citep{mullerSolarOrbiterMission2020}. 
Solar Orbiter follows an elliptical heliocentric orbit at varying distances from the Sun. Therefore, STIX observations are included only for events in which the angular separation between Solar Orbiter and the Aditya-L1 line of sight is less than \ang{10}, in order to minimise projection effects. The observations are corrected for light-travel time differences between the spacecraft.
Our sample is restricted to flares of class M and above that occur near the disk-centre ($\mu > 0.8$) and are not preceded by a flare of C-class or above in the same active region within the previous two hours. The study period spans July to November 2024, during the approach to solar maximum. Events satisfying these criteria during the study period are listed in Table~\ref{tab:flares_list}. The timing details of flares are based on the Heliophysics Events Knowledgebase\footnote{\href{https://www.lmsal.com/hek/}{https://www.lmsal.com/hek/}}. For comparison of source locations, we also use observations from the AIA and the Helioseismic and Magnetic Imager \citep[HMI;][]{Scherrer2012} onboard the Solar Dynamics Observatory \citep[SDO;][]{pesnell_solar_2012}.
    
\begin{table}
    \centering
    \resizebox{\linewidth}{!}{%
    \begin{tabular}[t]{lccccr}
    \hline
     No. & Flare & NOAA no.& Flare class &Flare start time\\

    \hline
    Case 1& SOL2024-07-10T05:59 &13738 &M1.5 &2024-07-10 05:44 \\
    Case 2& SOL2024-07-10T15:37 &13738 &M1.1 &2024-07-10 15:25 \\
    Case 3& SOL2024-10-09T01:56 &13848 &X1.8 &2024-10-09 01:25 \\
    Case 4& SOL2024-11-01T02:16 &13878 &M1.3 &2024-11-01 02:05 \\
    Case 5& SOL2024-11-01T14:31 &13878 &M2.0 &2024-11-01 14:18 \\
    Case 6& SOL2024-11-13T00:22 &13889 &M1.0 &2024-11-13 00:10 \\
    Case 7& SOL2024-11-13T17:08 &13889 &M1.7 &2024-11-13 16:57 \\
    
    \hline
    \end{tabular}}
    \caption{Flare cases analysed in this study. Flare class and start times are taken from the Heliophysics Events Knowledgebase (HEK).} 
    \label{tab:flares_list}
\end{table}

\section{Methodology}
\label{sec:methodology}
Transient brightenings during the pre-flare phase are detected in the \ion{Mg}{ii} h (2803{\AA}) and k (2796{\AA}) channels, with both filters identifying an identical set of events, given the similar formation heights of these two lines.  
For transient detection and localisation, \ion{Mg}{ii} h filter is used throughout this study. This choice is motivated by the higher availability of \ion{Mg}{ii} h ROI observations during flare localisation.
The analysis methodology described below is therefore based primarily on \ion{Mg}{ii} h  observations. 

Exposure time normalised light curves from the SUIT \ion{Mg}{ii} h images are generated. Careful inspection of these data reveals multiple small-scale brightenings during the pre-flare phase, which are localised using the base difference-image technique. 
The identified events are further examined using SUIT narrowband filters to estimate the transient signal at different atmospheric heights.
To obtain energy estimates for these transients, SUIT was cross-calibrated against IRIS observations.
These events are further analysed in conjunction with complementary X-ray coronal observations from SoLEXS,  HEL1OS, and STIX, whenever available. 
To establish the coronal counterparts of these chromospheric events and their relationship to the underlying magnetic field, SUIT observations are compared with HMI line-of-sight (LOS) magnetograms. Additionally, the presence of hot-onset conditions using both SoLEXS and STIX observations is compared with the chromospheric behaviour in the selected case studies.

\subsection{Definitions}
\label{sec:defs}
The definitions of key terms used in this paper are given below. 
\begin{itemize}
    \item \textbf{Excess intensity:}
Intensity enhancements within the ROI are estimated using a base-difference method. Pixels with intensities exceeding the detection threshold (described in \S\ref{subsec:identify transients}) are considered significant. The difference-image intensity of such features, measured relative to the base image, is defined as the excess intensity; the base difference image is considered as the background.

\item\textbf{Pre-flare transients:}
Local maxima identified in the excess-intensity light curve are defined as pre-flare transients. 
\item\textbf{Flaring region:}
\label{def:flare_regn}
The flaring region is defined as the set of pixels with intensity greater than 60\% of the peak intensity in the flare-peak image. 
\item\textbf{Flare impulsive phase:} The start time of the impulsive phase is defined as the time at which the HEL1OS or STIX light curve in the energy range $\geq$ 22 keV begins to show a systematic increase prior to the flare peak. This definition may slightly overestimate the true onset of the impulsive phase; however, it is a convention commonly adopted in hard X-ray studies \citep{battaglia_existence_2023}. The choice of 22 keV does not represent a physical threshold but reflects a practical criterion for isolating high-energy emission.
\item\textbf{Flare onset phase:}
The time interval from the GOES flare start time to the impulsive phase start time is considered the flare onset phase.
\item\textbf{Pre-flare phase:}
The time interval from two hours before the GOES peak time to the GOES flare start time is considered the flare pre-flare phase.

\end{itemize}
\subsection{Flare light curve}
All SUIT images are co-aligned to the temporally closest AIA 1600 {\AA}~ images using the \texttt{mapsequence\_coalign\_by\_match\_template} function, publicly available through the sunkit-image Python package\footnote{\href{https://docs.sunpy.org/projects/sunkit-image/en/v0.6.0}{https://docs.sunpy.org/projects/sunkit-image/en/v0.6.0}}, which is a Python recreation of a well-trusted and tested Solar Software (SSW) routine \texttt{tr\_get\_disp.pro} from the TRACE pipeline.

\begin{figure}
    \centering
    \includegraphics[trim={0cm 0.3cm 0cm 0.07cm}, clip, width=\linewidth]{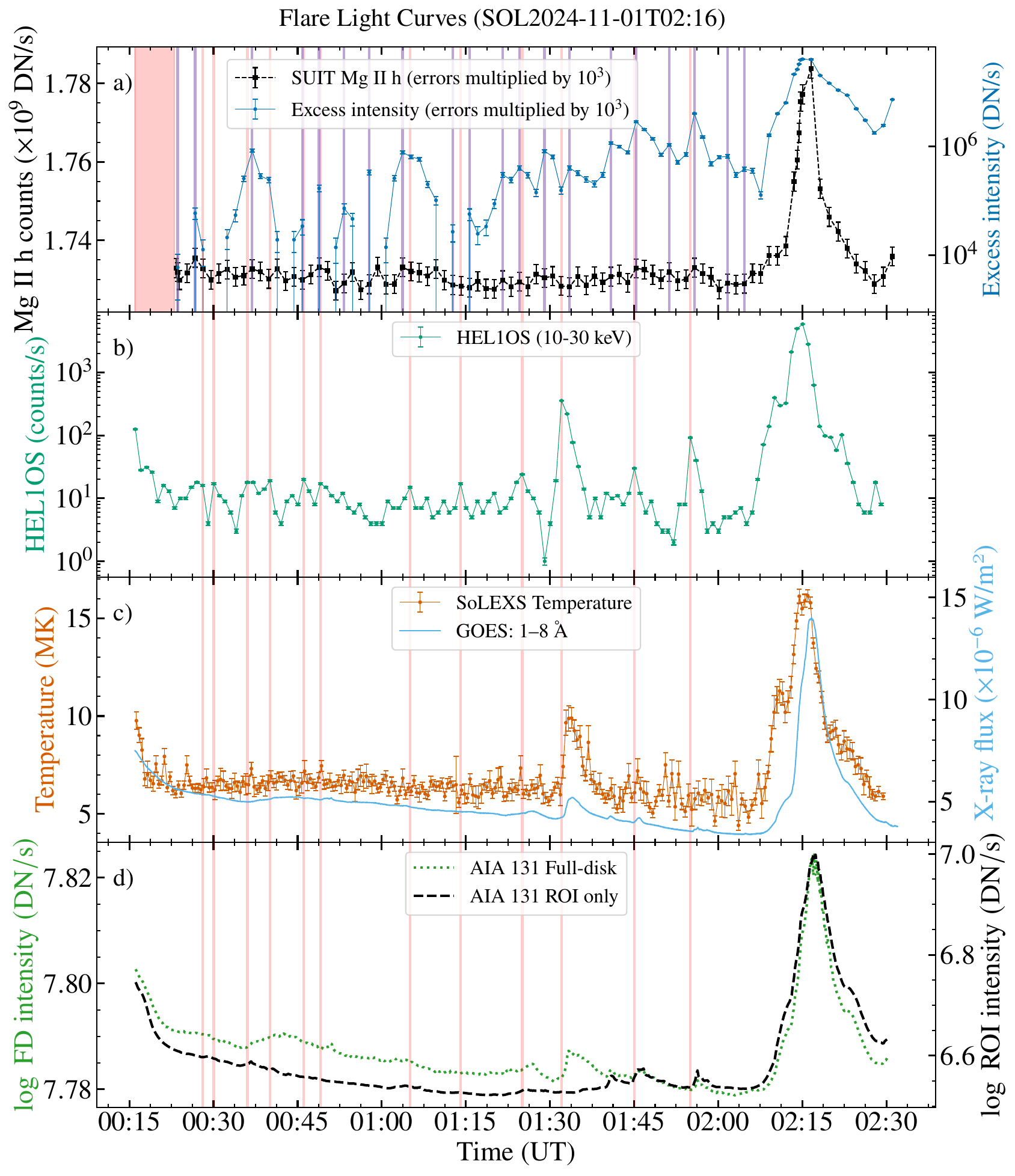}  
    \caption{a) Total intensity light curve of the \ion{Mg}{ii} h ROI images (black). The errors of the individual data points are scaled by $10^3$ for visibility. The light curve of the difference-image intensity above the threshold is shown in cyan, with errors scaled by $10^3$ for clarity. b) HEL1OS light curve (CdTe 1 and CdTe 2, 10{--}30~keV), binned to 1-minute intervals, c) Temperature derived from SoLEXS (orange) and X-ray flux from the GOES 1{--} 8 {\AA} band (sky blue), (d) Total intensity light curve from AIA 131 {\AA}~ integrated up to 1.1 \(R_\odot\) (green dot line), together with the AIA 131~{\AA} light curve extracted from the region corresponding to the SUIT ROI (black dashed line). The magenta vertical lines represent transients identified in \ion{Mg}{ii} h, while the red vertical lines represent transients identified from HEL1OS. The red shaded band indicates time intervals during which SUIT observations are unavailable.}
    \label{fig:c7_lc}
\end{figure}

This procedure is also used to correct for satellite drift between successive frames. The images are subsequently de-rotated to remove the effects of solar differential rotation. A common area is selected in all frames, and the total intensity within the area is computed. The resulting time series is used to construct the flare light curve. Poisson noise dominates over other sources of uncertainty; therefore, only Poisson errors are shown in the light curves (see Fig.\ref{fig:c7_lc} panel a).

\begin{figure*}
    \centering
    \includegraphics[width=0.9\linewidth]{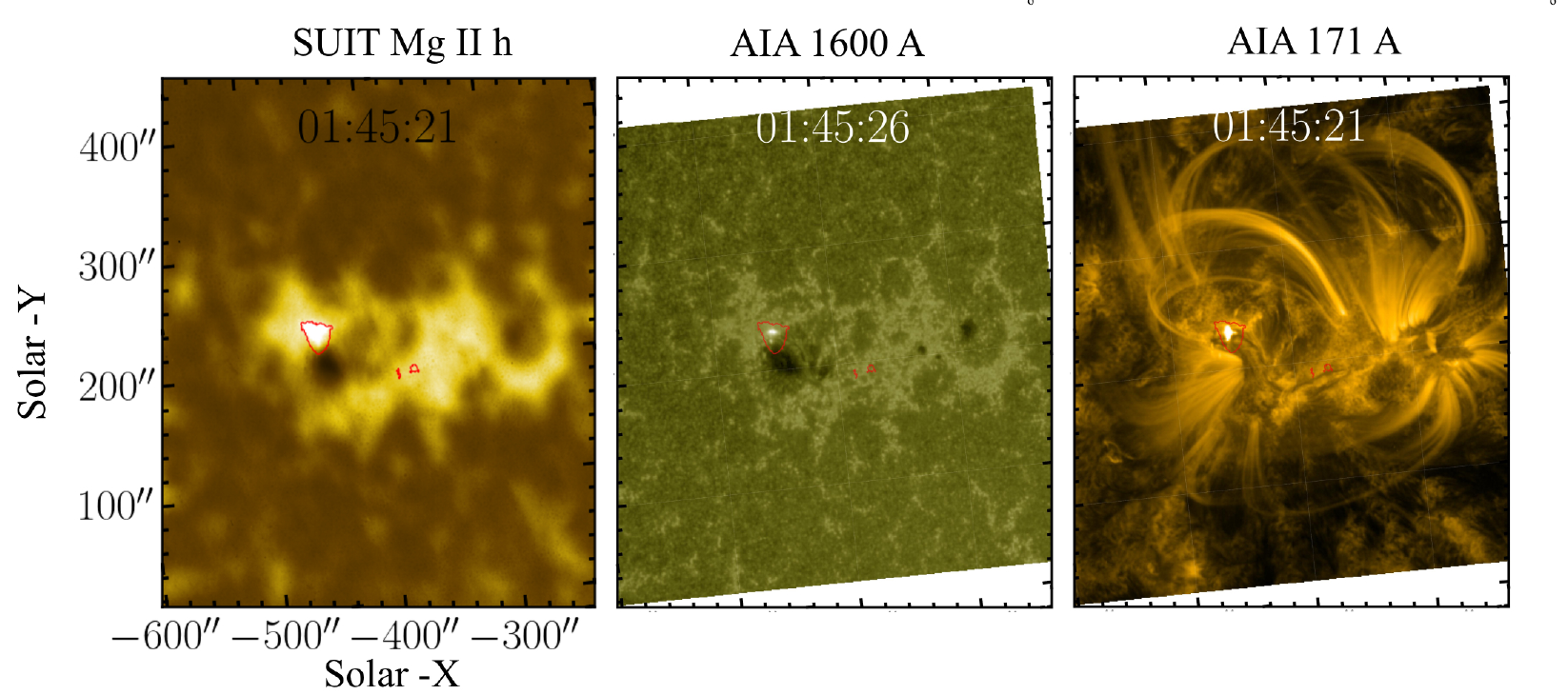}
    \caption{Example of a pre-flare transient identified in SUIT Mg II h observations, shown together with its counterparts in AIA 1600 {\AA}~ and AIA 171 {\AA}~ images. AIA images are tilted due to re-projection.}
    \label{fig:c7_compare_suit_aia_transient}
\end{figure*}

\begin{figure}
    \centering
    \includegraphics[width=0.9\linewidth]{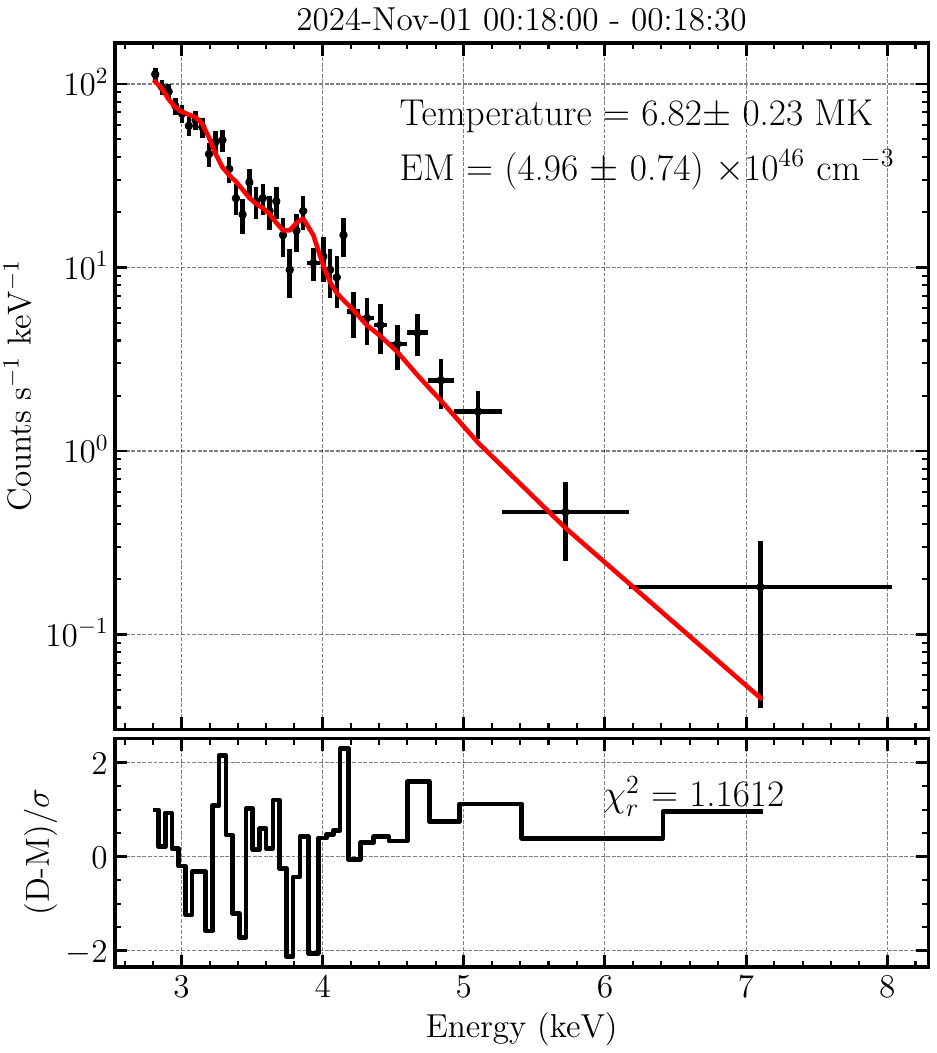}
    \caption{Representative isothermal spectral fit to the SoLEXS X-ray spectrum during the pre-flare phase. The observed SoLEXS count spectrum (black points) is fitted with an isothermal plasma model (red curve), assuming optically thin thermal bremsstrahlung and line emission. The lower panel shows the residuals between the observed spectrum and the model. The best fitting temperature and emission measure are $6.82\pm 0.23$ MK and  $(4.97\pm0.74)\times10^{46}cm^{−3}$, respectively.}
    \label{fig:solexs_fit}
\end{figure}

\begin{figure}
    \centering
    \includegraphics[trim={1cm 2.5cm 0cm 1cm}, clip, width=1.0\linewidth]{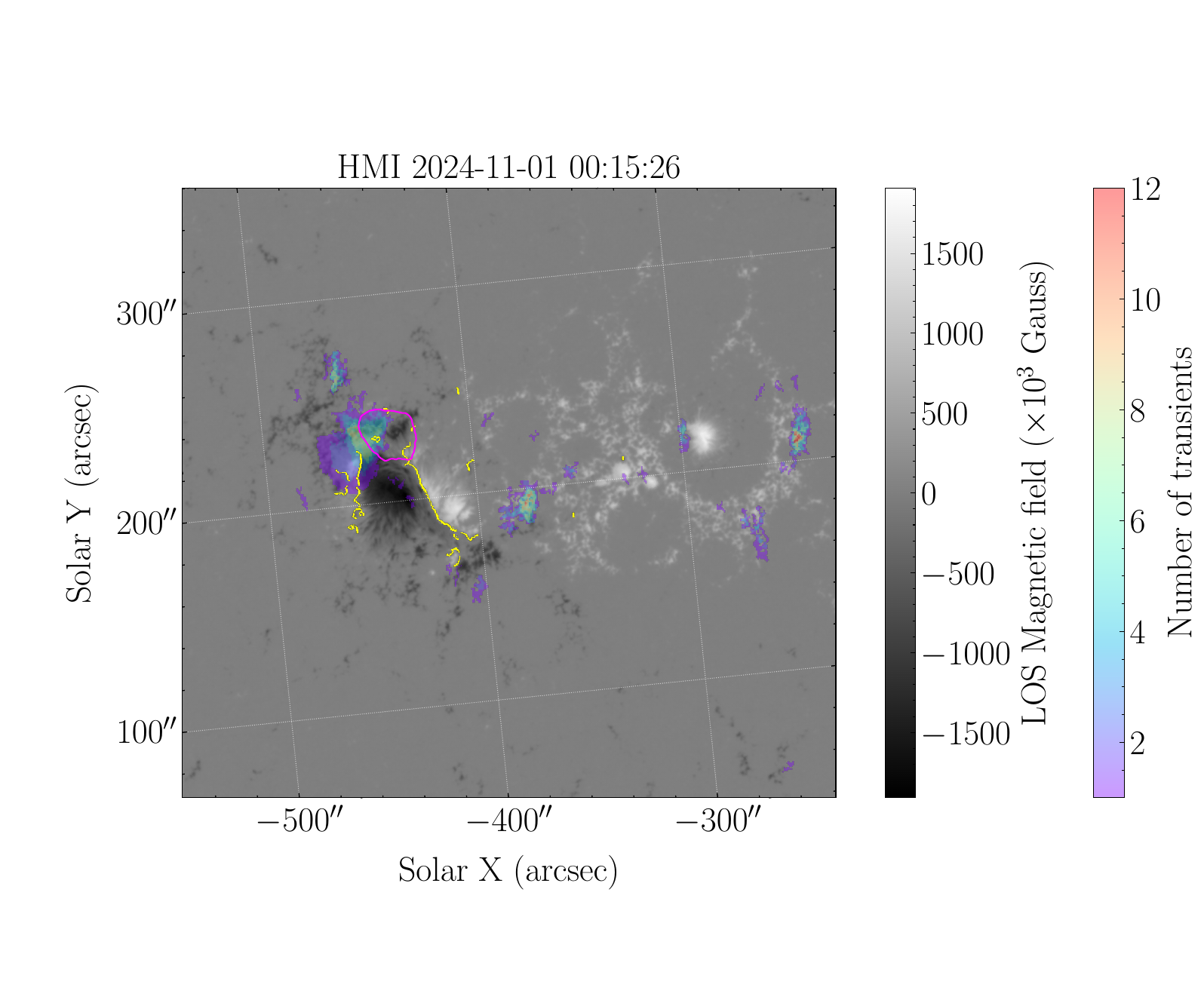}
    \caption{Case 4: All transient contours are stacked on the HMI magnetogram image using a rainbow colour scale. Strong magnetic polarity inversion lines are highlighted in yellow, and the magenta contour denotes the flaring region.}
    \label{fig:c4_hmi}
\end{figure}

\subsection{Identifying pre-flare transients in SUIT}
\label{subsec:identify transients}
The pre-flare transients are detected using the base-difference method. The base image was constructed from the median of the first five images in the 2-hour pre-flare image sequence, thereby minimising bias arising from selecting a single initial frame.
The base-difference images contain quiet-Sun and plage-related fluctuations, as well as intensity fluctuations caused by the contamination pattern. 
To mitigate these effects, a 5-sigma threshold is used to separate strong enhancements. 
Mean and standard deviation of the image were affected by bright events.
Therefore, we computed the median and standard deviation($\hat{\sigma}$) from the median absolute deviation (MAD) of the difference image, which is less affected by bright events. We define the detection threshold as,
\begin{align*} \label{eqn:threshold}
\text{Threshold} &= \tilde{X} + 5~ \hat{\sigma}\\ 
\hat{\sigma} &= k \times \mathrm{MAD}
\end{align*}

\noindent
where,
\begin{align*}
\tilde{X}    &= \mathrm{median}(X_i), ~~~~~~ \& ~~ \mathrm{MAD} = \mathrm{median}\left(|X_i - \tilde{X}|\right) \\
\mathrm{k} & \approx 1.4826 ~ \text{(for normal distribution)}
\end{align*}

Additionally, a minimum size constraint of 16 pixels (corresponding to twice the on-orbit PSF FWHM of $\approx$8 pixels) was imposed. 
One representative transient is shown in Fig. \ref{fig:c7_compare_suit_aia_transient}.
The base-difference total intensity within these selected regions was then used to generate the enhancement light curve.
All the local peaks in the excess intensity light curve are marked as pre-flare transient events (blue circles connected by a solid line, in Fig.\ref{fig:c7_lc}, panel a).
For each case, a list of all pre-flare transients within the 2-hour window from the peak flare time to the start of the flare is compiled.

\subsection{Cross calibration of SUIT with IRIS}
\label{sec:iris_eng_calib}
IRIS observations obtained on the same day were used for calibration in all cases, except Case~4, where IRIS coverage was limited to only part of the active region. In the remaining cases, IRIS and SUIT observed the same active region. 
Radiometric calibration of the IRIS rasters was performed using the \texttt{iris\_getwindata} routine along with the keywords \texttt{calib} and \texttt{perang} available in SSWIDL. 
The calibrated IRIS flux was integrated over the line profile within the full width at half maximum (FWHM) of \ion{Mg}{ii} h
and subsequently resampled to match the SUIT plate scale. 
An 8-pixel Gaussian smoothing was applied to approximate the SUIT point-spread function. The IRIS data were then co-aligned with the SUIT observations, and a scatter plot between IRIS flux and SUIT counts was used to derive the SUIT count-to-flux conversion factor. 
This calibration was applied to convert SUIT transient counts into flux values. 
To calculate the energy emitted at the source (luminosity), we assume the isotropic emission and multiply the flux value by $4\pi d^2$, where d is the distance between the Sun and the Earth.
Owing to the limited cadence of SUIT, the full event fluence is not reliably sampled; therefore, only the peak flux is considered in this analysis.
\subsection{HEL1OS light curves}
X-ray observations from HEL1OS are used to study the properties of the coronal hot plasma during and prior to flares. 
The HEL1OS 10–30 keV light curve was constructed by combining the event data from CdTe-1 and CdTe-2 detectors (see Fig. \ref{fig:c7_lc}, panel b). The error bars shown correspond to Poisson uncertainties. A median-filtered version with a 15-minute window was generated to identify transient features. Peaks standing above 3 $\sigma$ error bars are considered transient events in the X-ray light curve.
Multiple transients are observed in the 2-hour pre-flare window in the HEL1OS light curve; however, identifying the source regions of these transients is difficult, as HEL1OS observes the Sun as a star.

\subsection{Temperature from SoLEXS}
\label{subsec:Temperature_from_SoLEXS}
Solar flare spectra in the soft X-ray range are modelled using an isothermal emission model, with plasma temperature, emission measure, and elemental abundances as free parameters. The model spectrum is computed using the sunkit-spex\footnote{\href{https://github.com/sunpy/sunkit-spex}{https://github.com/sunpy/sunkit-spex}} Python package. 
The \texttt{f\_vth\_abun} model, which calculates the thermal continuum using optically thin bremsstrahlung emission and incorporates line emission based on the CHIANTI atomic database \citep[CHIANTI v7.1;][]{landiCHIANTIATOMICDATABASE2013} is employed, and the model is fitted to the observed spectrum using the Sherpa fitting application \citep[][]{siemiginowskaSherpaOpensourcePython2024,freemanSherpaMissionindependentData2001} to determine the best-fit parameters. 
During the pre-flare phase, elemental abundances are fixed to coronal values, whereas during the impulsive phase, the abundances are also allowed to vary. Fig.~\ref{fig:solexs_fit} presents a representative SoLEXS spectrum from the pre-flare phase, fitted with an isothermal model,
where the temperature and emission measure (EM) are treated as free parameters and the elemental abundances are fixed to coronal values.

\subsection{Identifying X-ray Source locations on full disc }
Soft and hard X-ray observations of the Sun are provided by the SoLEXS and HEL1OS; however, the source region of these activities cannot be determined from these observations. To locate the source of these enhancements, AIA observations at 131 {\AA}~ are used, which capture the hot plasma in the corona, and it is comparable to GOES 1{--}8 {\AA} among AIA channels \citep{fletcherFLARERIBBONENERGETICS2013}.
The light curves of the total counts full disc (up to 1.1 \(R_\odot\)), along with the active region corresponding to the SUIT field of view, are derived from AIA 131~{\AA} observations and plotted together.
The full-disk light curve is expected to mimic sun-as-a-star X-ray observations. If the selected active region is the source of the emission, a corresponding enhancement should also be visible in the active-region light curve. Conversely, if an enhancement is observed in the full-disk light curve but not in the active-region light curve, such enhancements are not attributed to the SUIT ROI.
Each enhancement/event in HEL1OS can be associated with a single active region (see the Fig.~\ref{fig:aia_131_all_lc}, which includes all light curves ).
These AIA light curves indicate which part of the X-ray light curve should be used to infer information about the active region being observed.
\subsection{Polarity inversion line (PIL)}
The polarity inversion line (PIL) is identified using the method described by \cite{ji_systematic_2023} using HMI LOS magnetograms (45s cadence). PILs are identified as the overlap between dilated strong-positive and strong-negative magnetic field regions (exceeding 100~G), and the masks are morphologically dilated to account for the finite separation between opposite polarities. 
Regions where the dilated opposite-polarity masks overlap are defined as strong-field PIL regions, and the PIL is obtained from the central line of the overlap (yellow colour line in Fig. \ref{fig:c4_hmi}).
\subsection{Spectral analysis of pre-flare enhancements}
\label{sec:spec_analysis}
\begin{figure}
    \centering
    \includegraphics[width=1\linewidth]{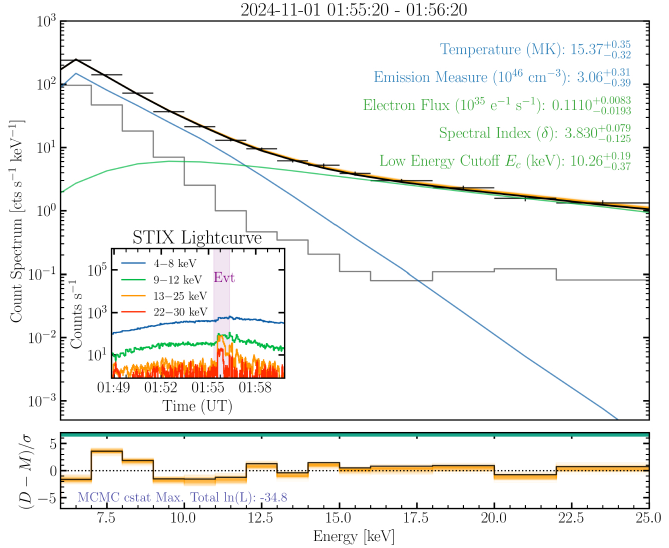}
    \caption{Pre-flare transient spectral fitting: Spectral fit was done for the data between 2024 November 01:55:20 to 01:56:20 UT. The observed count spectrum (black points) is fitted with a combined isothermal thermal component (\texttt{f\_vth}; blue curve) and a non-thermal component (\texttt{thick\_fn}; green curve), with the total model shown in black line. The lower panel shows the normalised residuals. The inset figure shows the STIX multi-energy-band light curve, with the magenta-shaded region indicating the time interval selected for spectral fitting. The best-fit parameters are listed in the figure legend.} 
    \label{fig:c7_spec_fitting}
\end{figure}

STIX onboard Solar Orbiter provides a substantially larger effective area compared to HEL1OS, enabling more robust spectral analysis.
However, since Solar Orbiter follows a heliocentric orbit, only events for which the STIX viewing angle relative to the Sun–Earth line is sufficiently small (less than \ang{10}) are considered to ensure consistency with Earth-based observations. Cases 3 to 7 satisfy these conditions.
To investigate the nature of pre-flare transients, X-ray spectra were fitted with both thermal and non-thermal emission models in the 6--25 keV energy range.
A representative example of a fitted spectrum is shown in Fig.~\ref{fig:c7_spec_fitting}.
Temperature measured by STIX is much higher than the temperature estimated from SoLEXS, which is discussed in \S\ref{sec:discussion}.
\subsection{Hot onset scenario}\label{subsec:hot_onset}
In most flare cases in this study, a pre-flare enhancement in \ion{Mg}{ii} h intensity was observed at the flaring location. Motivated by this behaviour, the presence of a hot X-ray onset, also known as a hot-onset precursor event (HOPE), was explored in these events. 
Hot-onset is characterised by  plasma temperatures 
already elevated above 10~MK at the earliest 
detectable point, accompanied by a monotonic increase in plasma emission measure in the absence of significant non-thermal emission prior to the impulsive phase.
To investigate the temporal evolution of thermal and nonthermal emission, we performed time-resolved X-ray spectral fitting of STIX data using \texttt{sunkit-spex}. 

\begin{figure}
    \centering
    \includegraphics[width=0.9\linewidth]{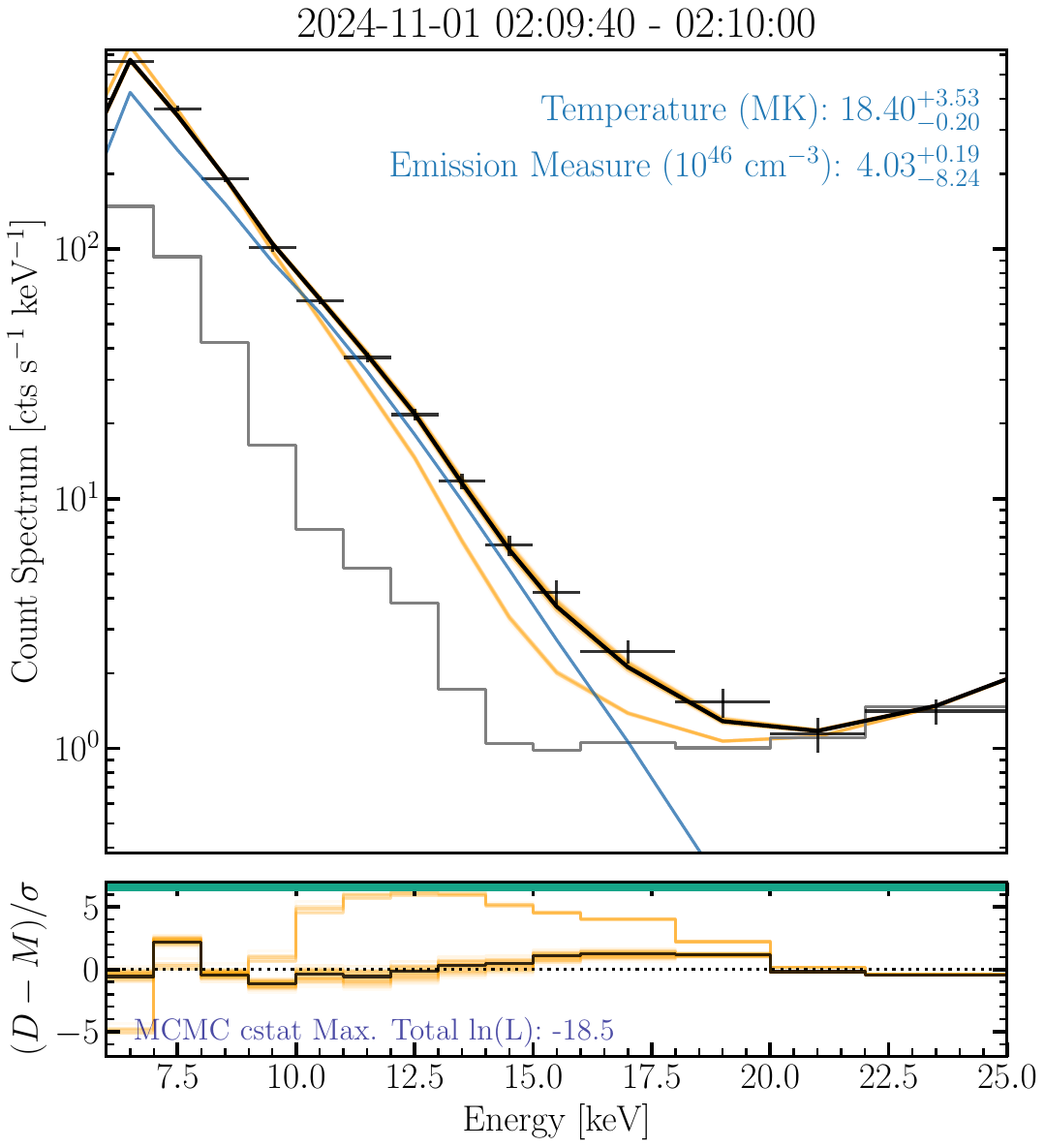}
    \caption{Representative STIX isothermal spectral fit during the flare onset phase. Black points denote the observed count spectrum, the solid black curve the best-fitting model, the blue curve the isothermal component, and the solid grey line the background derived from the quiescent interval (01:59 -- 02:03~UT on 2024 November 1).  The faint orange curves correspond to MCMC samples illustrating the uncertainty in the fitted model. The lower panel shows the normalized residuals, $(D-M)/\sigma$.}
    \label{fig:stix_hotonset_fit}
\end{figure}

For each event, a quiescent pre-event background spectrum was constructed by averaging over a 2--3~minute interval well before any detectable onset activity. Rather than directly subtracting the background counts, the background spectrum from this interval is incorporated 
into the spectral fitting as a scaled, simultaneous model component within \texttt{sunkit-spex}.The background model parameters are determined from the quiescent interval and subsequently held fixed during the fitting of the flare spectra (see Fig~\ref{fig:stix_hotonset_fit}). Spectra were extracted at 20-second intervals (increased to 30 seconds if no counts were detected). 

\begin{figure}
    \centering
    \includegraphics[width=0.8\linewidth]{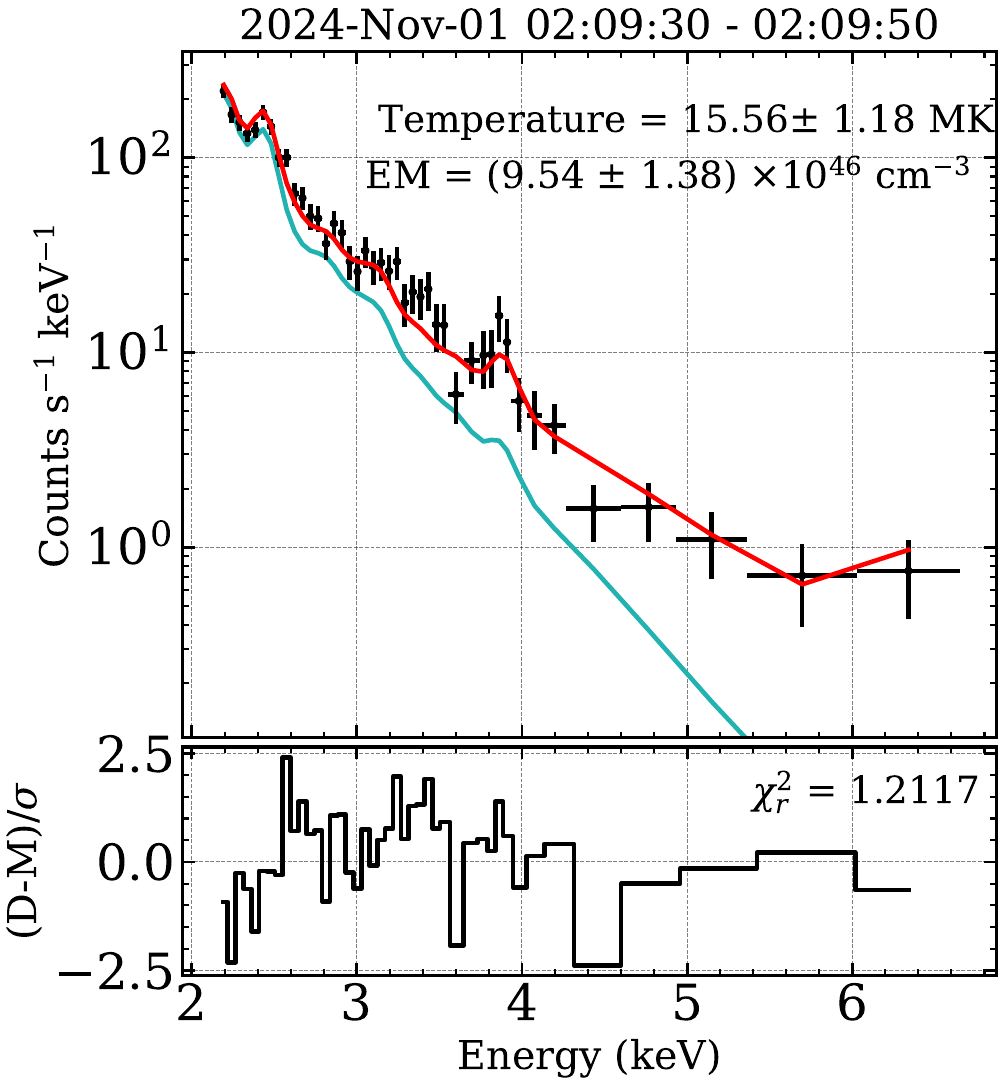   }
    \caption{Representative SoLEXS isothermal spectral fit during the flare onset phase. Black points denote the observed count spectrum, the solid red curve the best-fitting model, the cyan curve the background model derived from the quiescent interval (01:57 -- 02:00~UT on 2024 November 01). The lower panel shows the normalised residuals, $(D-M)/\sigma$.}
    \label{fig:solexs_hotonset_fit}
\end{figure}

\begin{figure}
    \centering
    \includegraphics[width=1\linewidth]{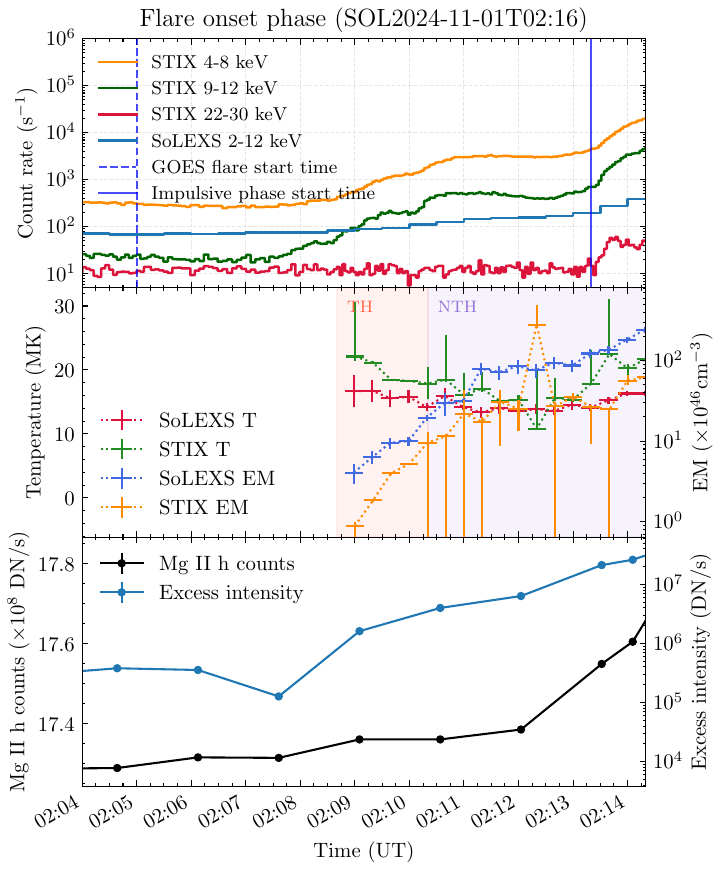} 
    \caption{Flare onset phase observed by STIX, SoLEXS and \ion{Mg}{ii} h. First panel light curves across different energy bands, with GOES flare start time (blue vertical dashed lines) and impulsive phase start time (blue solid vertical line) marked. The middle panel shows the temperature and emission measure derived from spectral fitting of STIX (green and orange) and SoLEXS (red and blue) spectra. The initial white region indicates the time interval for which reliable spectral fitting was not possible. The orange and purple-shaded background indicates the model used for the STIX spectral fit. The last panel shows the \ion{Mg}{ii} total count of the ROI and excess intensity, as in the first panel of Fig. \ref{fig:c7_lc}}
    \label{fig:hot_onset_check}
\end{figure}

Each spectrum was fitted over the 6{-–}25 keV energy range using two model combinations: (i) an isothermal component (\texttt{f\_vth}) alone, and (ii) an isothermal component combined with a collisional thick-target bremsstrahlung model \citep[\texttt{thick\_fn};][]{brownDeductionEnergySpectra1971,holmanImplicationsXrayObservations2011}. Model selection was performed using the maximum log-likelihood (ln(L)) from MCMC sampling. 
A difference $\Delta$ ln(L) > 10 between the two-component and single-component fits was adopted as the threshold for a statistically significant non-thermal detection. This constitutes strong evidence in favour of the more complex model, and corresponds to a p-value of $\approx 1.7 \times 10^{-4} (\sim  3.7 \sigma)$ under a likelihood ratio test with 3 degrees of freedom.
The free parameters of the thick-target model —total electron flux, spectral index $\delta$, and low-energy cutoff $E_c$ — were left unconstrained during fitting; however, results with $\delta$ > 8 or $E_c$ < 5 keV were flagged as physically unreliable due to degeneracy with the thermal component. 
An analogous background subtraction was applied to SoLEXS spectra during the onset phase to derive
the corresponding temperature and emission measure evolution (see Fig.~\ref{fig:solexs_hotonset_fit}); the SoLEXS fitting procedure is described in 
Section \S\ref{subsec:Temperature_from_SoLEXS}.
A representative plot of the hot-onset scenario of case 4 is presented in Fig. \ref{fig:hot_onset_check}. 

\section{Results}
\label{sec:results}
Analysis of the seven events listed in Table~\ref{tab:flares_list} revealed a total of 102 pre-flare transients in the \ion{Mg}{ii} h observations. The transients detected in the \ion{Mg}{ii} k filter are identical to those observed in \ion{Mg}{ii} h, consistent with the similar formation heights of these two lines. In contrast, no transient events are detected in the continuum filters of the SUIT, like 2767 {\AA}, 2832 {\AA}, 3000{\AA}, and 3880 {\AA}.
In this section, two representative cases of the pre-flare phase are presented. These two cases are selected based on the availability of STIX data and their distinct observational characteristics. All the cases are summarised in Table~\ref{tab:summary_table}.

\subsection{M1.3 class flare on 2024  November 1}
The M1.3-class flare associated with NOAA AR 13878 on 2024 November 1, began at 02:05 UT and reached its peak at 02:16 UT 
in the GOES 1–8 {\AA}~ band. Fig.~\ref{fig:c7_lc} shows the light curve of the total intensity within the ROI in the \ion{Mg}{ii} h filter (top panel, black squares connected by dashed lines), covering a time interval from two hours before the flare peak to 15 minutes after the peak. 
Excess intensity, as defined in the \S(\ref{sec:defs}), is also plotted in the same window, and shown as blue circles connected by a solid line. Local peaks of the excess intensity light curve are identified and marked as transients. 
20 such transient brightenings were identified during the pre-flare phase in this case (see Fig. \ref{fig:c7_lc}).
Panel (b) shows the 10{--}30 keV light curve from HEL1OS, obtained by combining data from both CdTe 1 and CdTe 2 detectors and binned at 1-minute intervals (green circles connected by a solid line). The transients identified in the HEL1OS light curve are marked with red vertical lines and extended to panel (a) to investigate their co-temporal behaviour.
The temperature derived from SoLEXS for the same observation period is shown in panel (c) (orange circles connected by a solid line). The GOES 1–8~{\AA} light curve (cyan solid line) is also included for comparison with the flare profile. Light curves derived from AIA 131~{\AA}, for both the full disc (green dotted line) and the ROI (black dashed line), are shown in panel (d). 

Images of one representative transient, in different atmospheric heights, are shown in  Fig.~\ref{fig:c7_compare_suit_aia_transient}.
The SUIT Mg II h image (left) contains the identified transient marked with a red contour. The same contour is overplotted on the AIA 1600 {\AA} (middle) and the AIA 171 {\AA} (right).
The tilt in the AIA images is due to re-projection to match the SUIT ROI.
The transient was observed in both channels and co-spatial over different heights from the upper photosphere to the corona. 
The transient contours identified in SUIT appear slightly larger than their counterparts in the AIA 1600 {\AA} and AIA 171 {\AA} observations, which is expected given the broader point-spread function (PSF) of SUIT. 
A detailed timing comparison with other instruments is limited by SUIT's relatively low cadence ($\approx 90s$) in this case.
All transient contours are stacked on the first-frame HMI line-of-sight (LOS) magnetogram to examine the spatial distribution of pre-flare activity and its association with the underlying photospheric magnetic field.

In Fig.~\ref{fig:c4_hmi}, the colour scale represents the number of overlapping transient contours at each location. Warmer colours indicate a higher occurrence frequency, with red corresponding to locations where transients are most frequently observed.
The magenta contour in the image represents the location of the flare (described in \S\ref{def:flare_regn}). It is observed that transients are more concentrated in the midst of the flaring region. A large fraction of the transients are co-spatial with the PIL, while a smaller number are located away from it (see Fig. \ref{fig:c4_hmi}). 

Several of the chromospheric enhancements exhibit corresponding pre-event signatures in HEL1OS. Ten HEL1OS peaks have clear SUIT counterparts. However, not all SUIT transients exhibit associated HEL1OS peaks. At 01:39 UT, HEL1OS shows a distinct peak, whereas the corresponding enhancement in SUIT  showed a small peak (see Fig. \ref{fig:c7_lc}). Further investigation using AIA 131 {\AA}~ data reveals that a small flare occurred in a different active region at this time (see Fig. \ref{fig:c7_lc}, panel d). The active region in this ROI also shows a small brightening in response to the event, suggesting a magnetic or energetic connection between the two active regions.

STIX observations are available for this event, including the pre-flare phase. All co-temporal HEL1OS peaks with SUIT are fitted with a thermal and a nonthermal model (as explained in \S\ref{sec:spec_analysis}). The spectra taken during the pre-flare phase at 01:45 UT and 01:55 UT have counts in higher-energy ranges that are significant enough to achieve a good fit of the data with the combined thermal and non-thermal model (\texttt{f\_vth + thick\_fn})(see Fig. \ref{fig:c7_spec_fitting}), remaining transients are fitted well with a thermal-only model. 
A continuous enhancement in \ion{Mg}{ii} h intensity begins to appear at one of the footpoints, away from the flaring location. After 01:43 UT, approximately 47 minutes prior to the soft X-ray peak time derived from GOES observations, and disappeared after 02:07 (GOES flare start time is 02:05). 
During this interval, multiple small-scale events are detected in HEL1OS. The locations of all these events coincide with the same active region, as confirmed by AIA 131~{\AA} observations.
The plasma temperature derived from SoLEXS remains in the range of 6–8 MK during the pre-flare phase and shows no significant variation associated with the transients (see Fig. \ref{fig:c7_lc} panel c).

Figure~\ref{fig:hot_onset_check} shows the flare onset phase. The top panel shows STIX and SoLEXS light curves across different energy bands, with the GOES flare start time (blue dashed vertical line) and the impulsive phase start time (blue solid vertical line) marked. The 4–8 keV band begins to rise at 02:07 UT, approximately 6 minutes before the impulsive phase (02:13 UT). Spectral fitting of the background-modelled STIX and SoLEXS spectra yields reliable fits from 02:08:40~UT onward, when the background-subtracted signal-to-noise is sufficient for stable parameter estimation (SNR above 3$\sigma$ in 6 to 12 keV) energy range. The temperature and emission measure derived from SoLEXS (red and blue) and STIX (green and orange) are shown in the middle panel. Note that during the interval 02:08:40--02:10:20~UT, only the isothermal model is used (see Fig. \ref{fig:stix_hotonset_fit} and Fig. \ref{fig:solexs_hotonset_fit}). Adding a non-thermal component to STIX spectra does not significantly improve the fit ($\Delta \ln L < 10$), indicating that the emission is predominantly thermal during this phase.
SoLEXS is sensitive primarily to thermal plasma; it cannot constrain the presence of non-thermal emission. The temperatures derived from SoLEXS remain $\approx$ 15 MK, while STIX yields temperatures of approximately 15{--}20 MK, which are systematically slightly higher than the SoLEXS values. Nevertheless, the two different temperatures, both above 10 MK, measured by two instruments remain nearly constant during the onset phase, while the emission measure increases monotonically between 02:08:40 and 02:10:20~UT, satisfying the hot-onset condition. As the soft X-ray emission begins to rise, the \ion{Mg}{ii} h excess intensity also shows a gradual increase, followed by a rapid enhancement after the onset of the impulsive phase, at the flaring location (see Fig. \ref{fig:hot_onset_check}).

\begin{figure}
    \centering
    \includegraphics[width=1\linewidth]{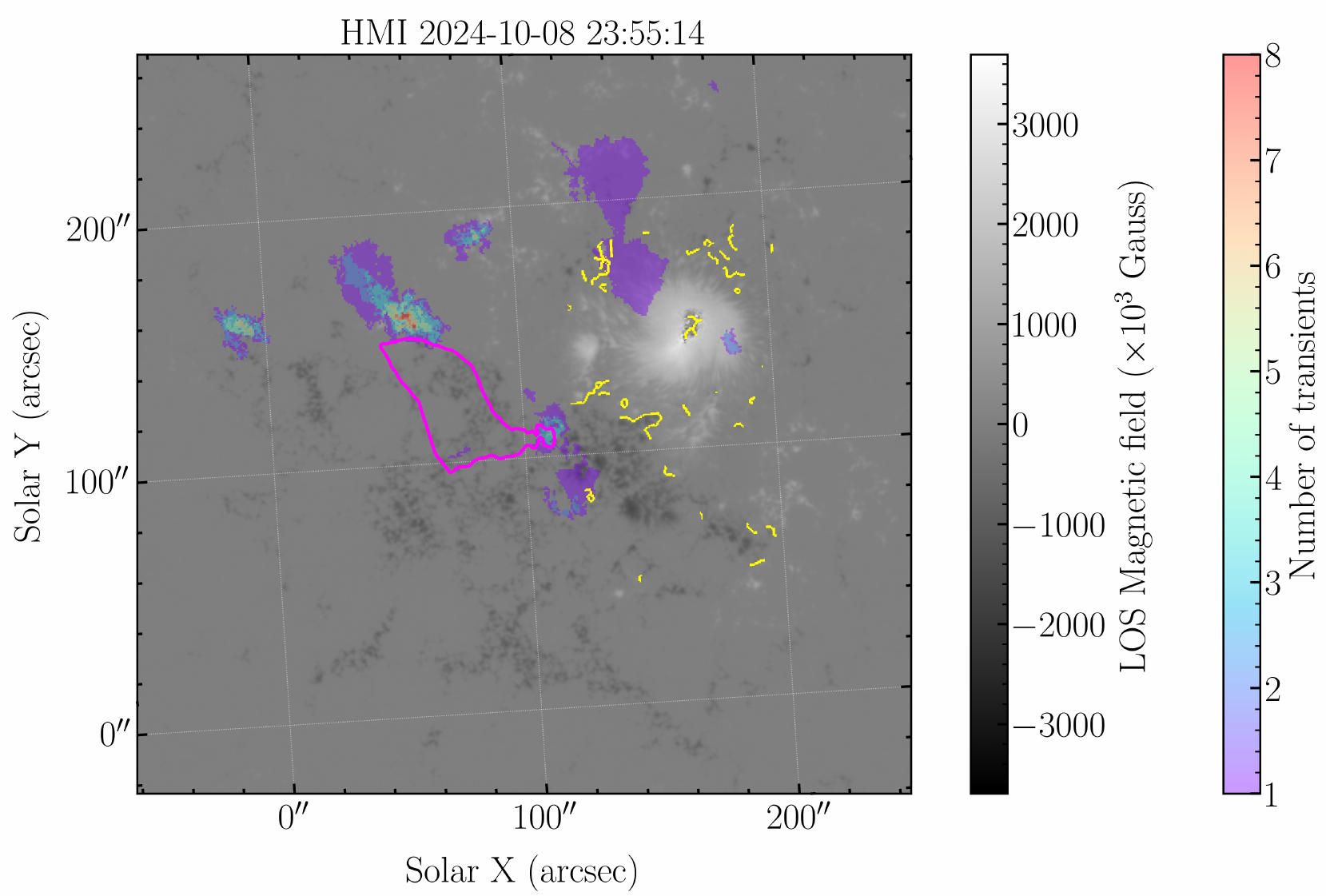}
    \caption{Case 3: All transient contours are stacked on the HMI magnetogram image using a rainbow colour scale. Strong magnetic polarity inversion lines are highlighted in yellow, and the magenta contour denotes the flaring region.}
    \label{fig:c6_hmi}
\end{figure}
\subsection{X1.8 class flare on 2024 October 9 }
\begin{figure}
    \centering
    \includegraphics[width=1\linewidth]{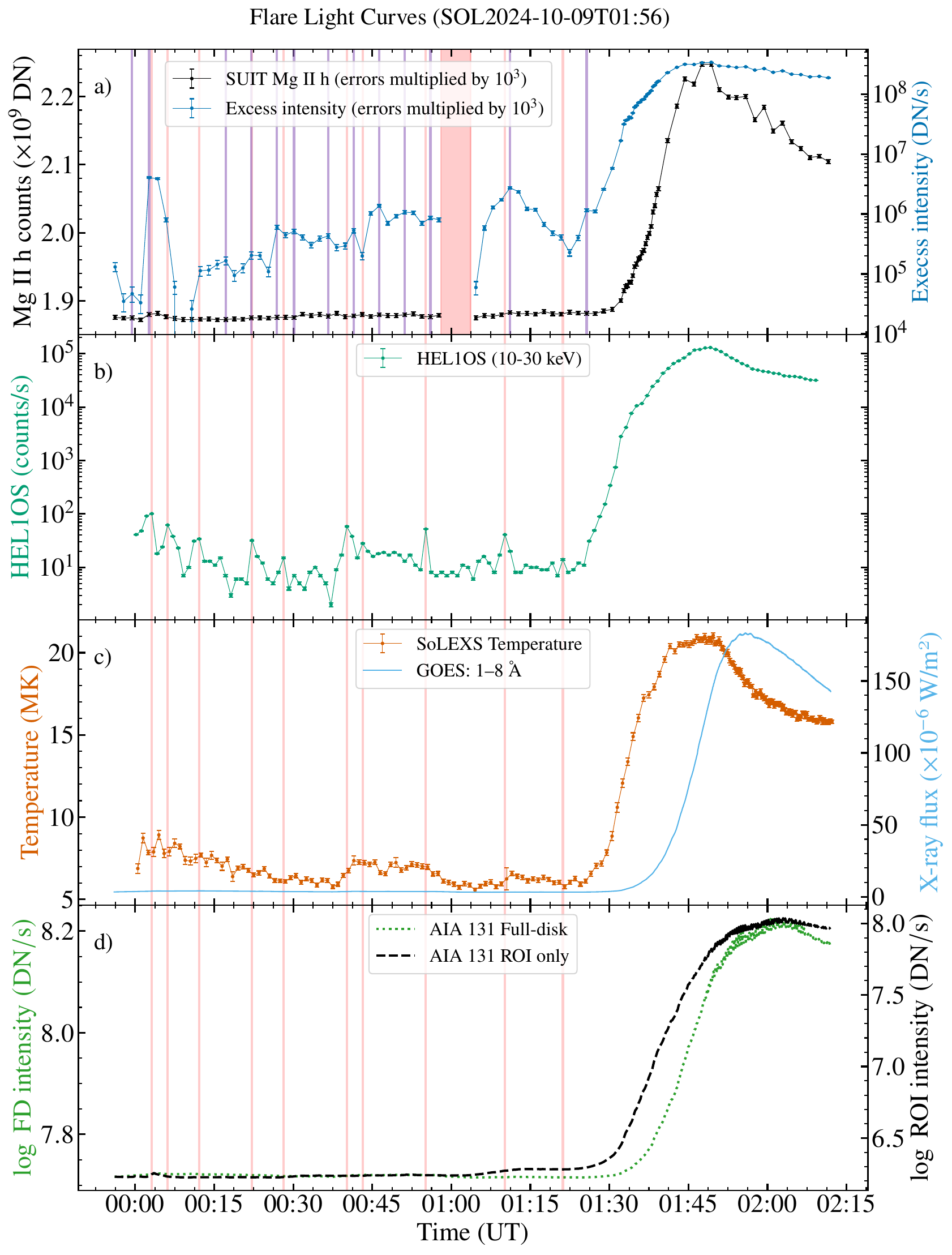}
    \caption{a) Total intensity light curve of the \ion{Mg}{ii} h ROI images (black). The errors of the individual data points are scaled by $10^3$ for visibility. The light curve of the difference-image intensity above the threshold is shown in cyan, with errors scaled by $10^3$ for clarity. b) HEL1OS light curve (CdTe 1 and CdTe 2, 10{--}30 keV), binned to 1-minute intervals, c) Temperature derived from SoLEXS(orange) and X-ray flux from the GOES 1{--} 8 {\AA} band (sky blue), (d) Total intensity light curve from AIA 131 {\AA}~ integrated up to 1.1 \(R_\odot\) (green dot line), together with the AIA 131~{\AA} light curve extracted from the region corresponding to the SUIT ROI (black dashed line). The magenta vertical lines represent transients identified in \ion{Mg}{ii} h, while the red vertical lines represent transients identified from HEL1OS. The red shaded band indicates time intervals during which SUIT observations are unavailable.}
    \label{fig:c6_lc}
\end{figure}

\begin{figure}
    \centering
    \includegraphics[width=1\linewidth]{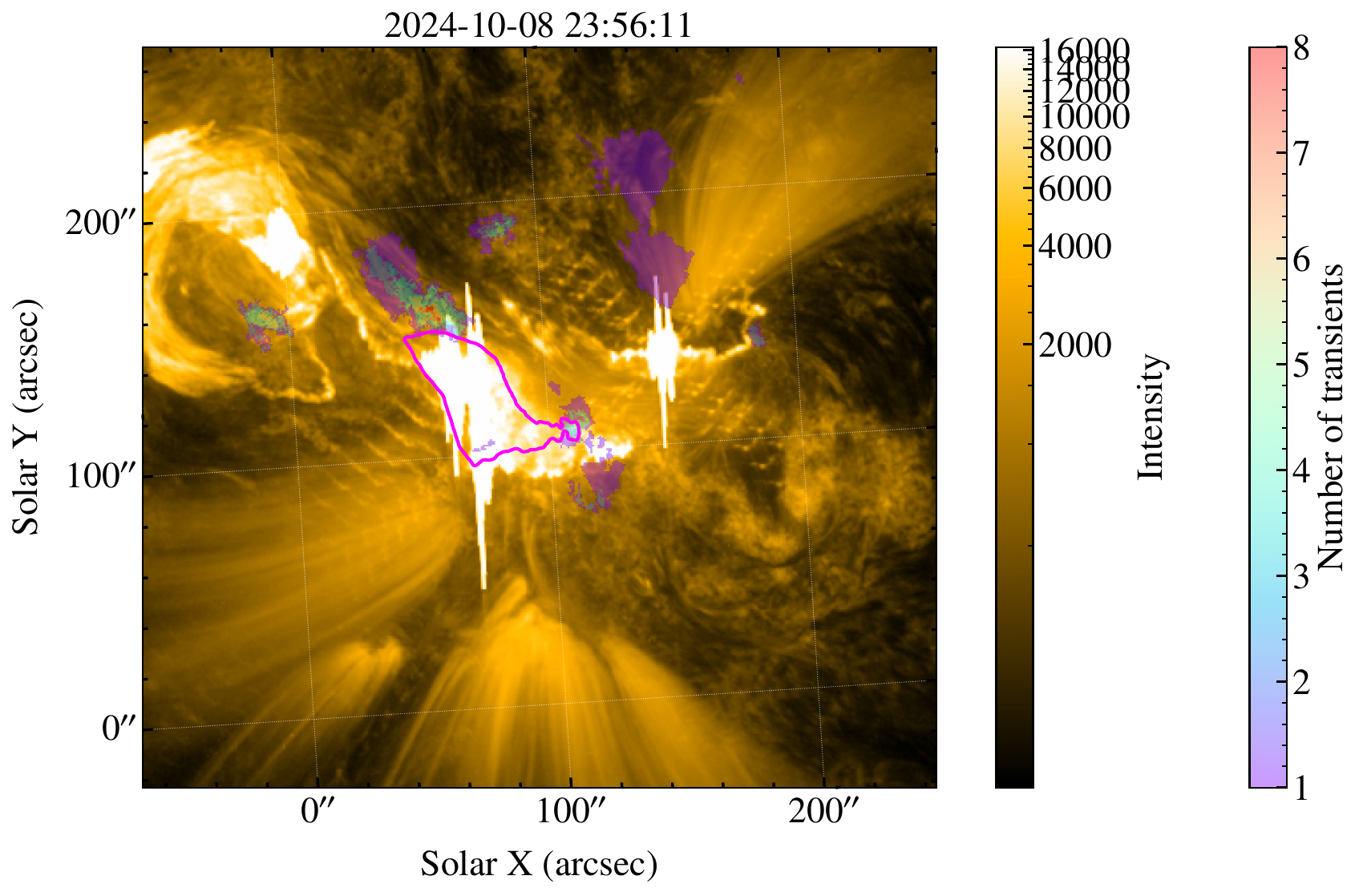}
    \caption{Case 3: All transient contours are stacked on the AIA 171 {\AA} image (corresponding to \ion{Mg}{ii} h peak intensity time, de-rotated to start time) using a rainbow colour scale. The magenta contour denotes the flaring region.}
    \label{fig:c6_171}
\end{figure}

The X1.8-class flare associated with NOAA AR 13848 on 2024 October 9 began at 01:25 UT and reached its peak at 01:56 UT in the GOES 1–8 Å band (see Fig. \ref{fig:c6_lc}).
This event is the strongest flare in our dataset and exhibits a sigmoid morphology  \citep[see ][]{rust_evidence_1996, Gibson2002}.
We identified 13 pre-flare transients, and 5 of them had HEL1OS  counterparts.
The pre-flare transients detected in UV using SUIT, during this event, do not spatially overlap with either the flare ribbon contours or the PIL (see Fig. \ref{fig:c6_hmi}); they are spatially associated with the sigmoid loop structure observed in AIA 171 {\AA}~ images (see Fig.~ \ref{fig:c6_171}). Two prominent pre-flare brightenings are observed at approximately 00:04 UT and 01:12 UT. Both are associated with enhanced emissions originating from different active regions, as evident from AIA observations. Simultaneously, the target active region shows activity in AIA 171~{\AA}, which is reflected as a corresponding response in the \ion{Mg}{ii} h channel.

\begin{figure}
    \centering
    \includegraphics[width=1\linewidth]{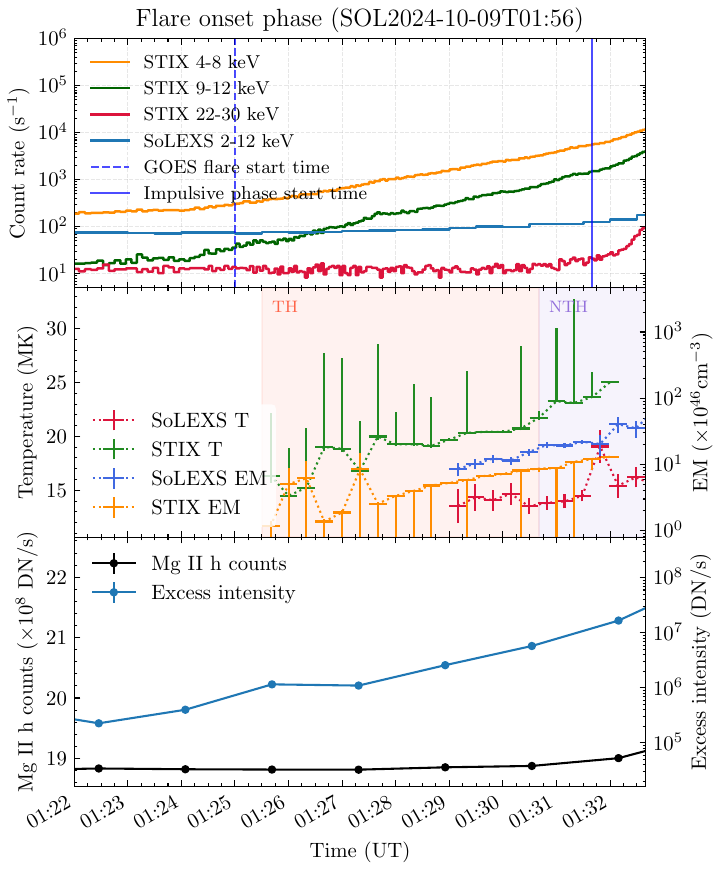}
    \caption{Flare onset phase observed by STIX, SoLEXS and \ion{Mg}{ii} h. First panel light curves across different energy bands, with GOES flare start time (blue vertical dashed lines) and impulsive phase start time (blue solid vertical line) marked. The middle panel shows the temperature and emission measure derived from spectral fitting of STIX (green and orange) and SoLEXS (red and blue) spectra. The initial white region indicates the time interval for which reliable spectral fitting was not possible. The orange and purple-shaded background indicates the model used for the STIX spectral fit. The last panel shows the \ion{Mg}{ii} total count of the ROI and excess intensity, as in the first panel of Fig. \ref{fig:c6_lc}}
    \label{fig:c6_stix_hot_onset}
\end{figure}

The X-ray light curve during the pre-flare phase includes contributions from multiple active regions. In particular, the enhancement between 00:30 and 01:00 UT is dominated by emission from a separate active region (NOAA regions 13849 and 13850), as confirmed by AIA light curves, while the target region (NOAA region 13852) contributes at other times. This highlights the importance of spatial discrimination when interpreting Sun-as-a-star X-ray observations.
During the onset phase, all X-ray bands exhibit a monotonic increase (see Fig.\ref{fig:c6_stix_hot_onset}). 
Reliable spectral fitting is achieved from 01:25:40 UT for STIX and from 01:29:00 UT for SoLEXS. As in the previous case, the temperatures derived from STIX are systematically higher than those estimated from SoLEXS. The hot-onset condition is satisfied during the interval 01:29--01:31 UT. The non-thermal component becomes significant close to the impulsive phase. 
We also observe a small offset between the temperature peak and the soft X-ray light curve, with the temperature reaching its maximum slightly before the soft X-ray peak.
The excess intensity in \ion{Mg}{ii} h follows the trend of the low-energy X-ray emission and shows a rapid increase after the impulsive phase. 

\subsection{Statistical properties of pre-flare transients}
The peak radiated energies of all 102 transient events identified in the seven pre-flare cases are analysed. SUIT provides flux measurements in units of DN~s$^{-1}$; these are converted to physical units (erg~s$^{-1}$~cm$^{-2}$) through cross-calibration with IRIS, as described in Section~\ref{sec:iris_eng_calib}. The flux of each transient is then estimated by integrating the radiated energy within the selected contour. To investigate the statistical properties of the pre-flare transients, we construct the complementary cumulative distribution function (CCDF: $P(X \geq x)$) of their peak time flux. Fig.~\ref{fig:ccdf_mle} shows the CCDF of all selected transients (black pluses).
Of the 49 detected X-ray counterparts, 29 are associated with the active region under study. These events are indicated by red circles.
The measured peak flux of SUIT transients varies from 0.01 $erg~cm^{-2}~s^{-1}$ (Luminosity $= 3.6 \times 10^{25} erg~s^{-1}$ ) to 2.9 $erg~cm^{-2}~s^{-1}$ (Luminosity $=8.04 \times 10^{27} erg~s^{-1}$ ). 

We estimate the power-law behaviour of the distribution using maximum likelihood estimation (MLE), following the approach of \cite{clausetPowerLawDistributionsEmpirical2009} and \cite{milligan_lyman-alpha_2020}. MLE is preferred for relatively small datasets, as it is not sensitive to binning choices \citep{dhuysEffectLimitedSample2016}. Based on the observed distribution, we identify two break energies and divide the distribution in Fig.\ref{fig:ccdf_mle} into 3 segments (representative images from each segment are shown in Fig. \ref{fig:3type}). 
The first segment contains 35 events and is almost flat on the CCDF. Since there are no X-ray counterparts for these events, they are either weak responses to nearby coronal activity, or could be very faint flare-like events where energy estimation is not reliable due to higher background plage brightness (see left panel Figure \ref{fig:3type}).

The transients in the second and third segments exhibit significant UV emission and appear as small flares. 
We fit the truncated MLE for the second and third segments \citep{delucaFittingGoodnessoffitTest2013a} by maximising the log-likelihood function.

\begin{equation*}
  f(x \mid \alpha) =
  \frac{1 - \alpha}{x_{\max}^{1-\alpha} - x_{\min}^{1-\alpha}}\,
  x^{-\alpha},
  \quad x_{\min} \leq x \leq x_{\max},
  \label{eq:truncpl}
\end{equation*}
with log-likelihood
\begin{equation*}
  \ln \mathcal{L}(\alpha) =
  N\ln\!\left(
    \frac{1-\alpha}{x_{\max}^{1-\alpha}-x_{\min}^{1-\alpha}}
  \right)
  - \alpha \sum_{i=1}^{N} \ln x_i.
  \label{eq:logl}
\end{equation*}
The resulting power-law indices below and above the break are 1.64 and 3.12, respectively (see Fig.~\ref{fig:ccdf_mle}). 
We note that, for the CCDF, the slope is related to the differential power-law index as $\alpha_{CCDF}=\alpha−1$, and this relation has been accounted for when plotting the CCDF \citep{clausetPowerLawDistributionsEmpirical2009}.
\begin{figure}
    \centering
    \includegraphics[width=1\linewidth]{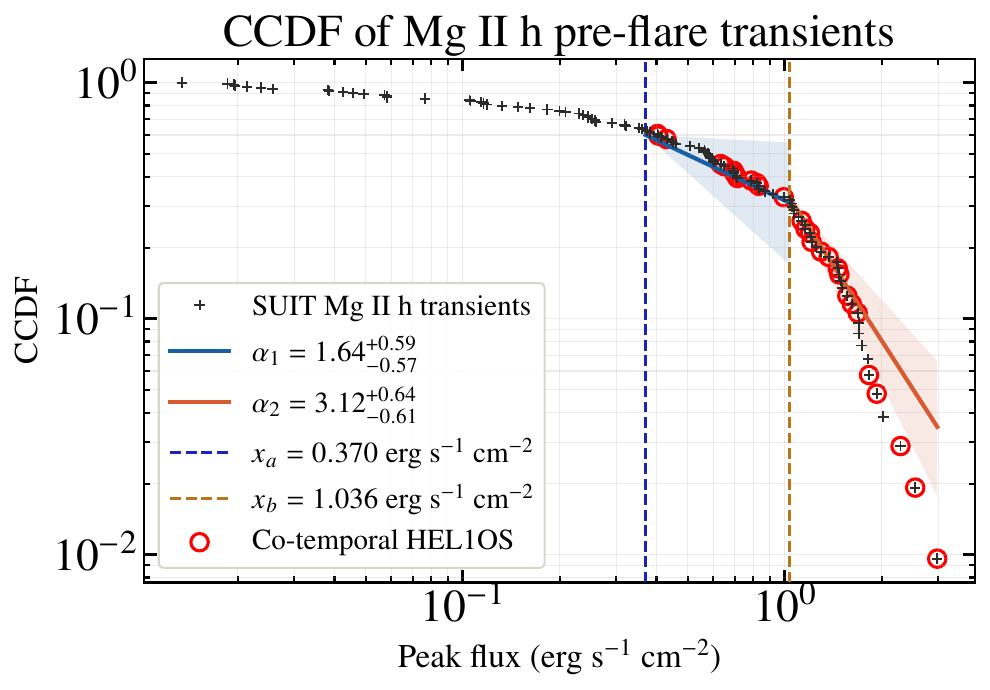}
    \caption{Complementary cumulative distribution function (CCDF) of fully detected pre-flare transients (black pluses). Events with X-ray counterparts observed by HEL1OS, originating from the same active region as the SUIT ROI, are marked with red circles. The CCDF is overplotted with maximum-likelihood estimated (MLE) power-law models (blue and orange lines). The vertical blue dashed line indicates the lower cut-off applied, while the orange dashed line marks the break energy.}
    \label{fig:ccdf_mle}
\end{figure}
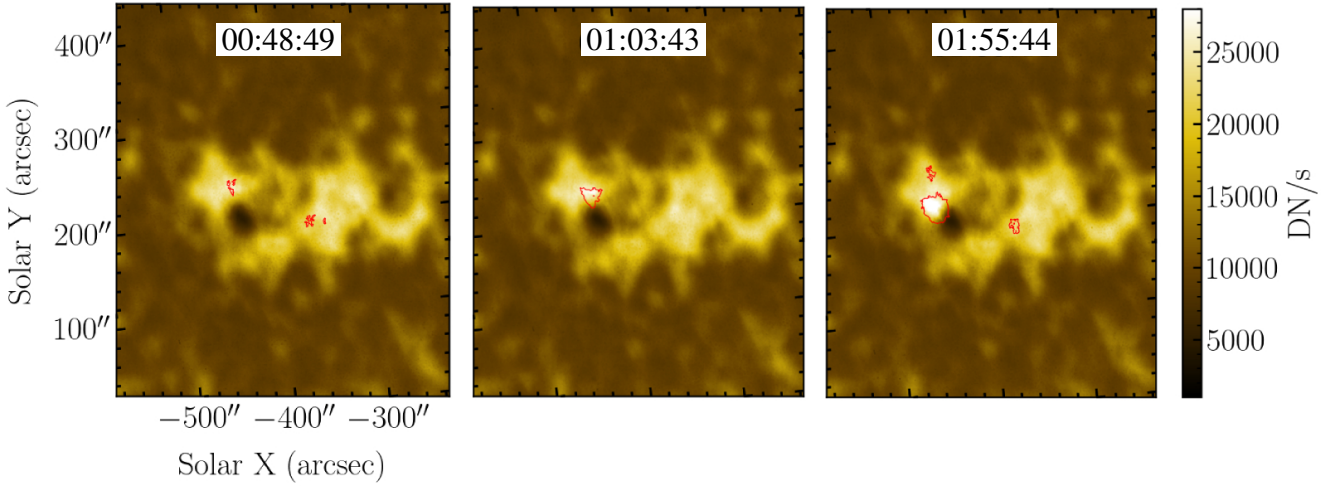
\begin{figure*}
    \centering
    \begin{tikzpicture}
        \node[anchor=south west,inner sep=0] (img) at (0,0)
            {\includegraphics[width=\textwidth]{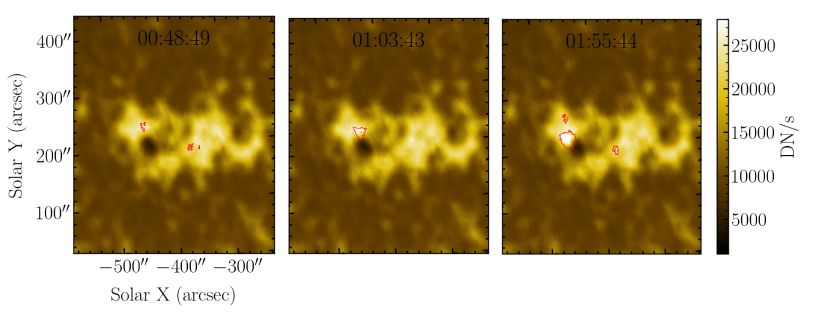}};
        \begin{scope}[x={(img.south east)}, y={(img.north west)}]
            \node[fill=white, inner sep=2pt,font=\Large] at (0.21,0.88) {00:48:49};
            \node[fill=white, inner sep=2pt,font=\Large] at (0.48,0.88) {01:03:43};
            \node[fill=white, inner sep=2pt,font=\Large] at (0.74,0.88) {01:55:44};
        \end{scope}
    \end{tikzpicture}
    \caption{Representative transient events from three segments of Fig.~8 observed on 2024 November 1. The left panel shows a small event from the first segment, the middle panel corresponds to the second segment, and the right panel shows an event from the third segment.}
    \label{fig:3type}
\end{figure*}

\section{Summary and Discussion}
\label{sec:discussion}
In this paper, we investigate the pre-flare phase of multiple solar flares using a unique set of observations from Aditya-L1, the latest multi-instrument observatory-class satellite mission dedicated to solar studies. During the pre-flare phase, we identify multiple short-lived chromospheric transients in the \ion{Mg}{ii} h and k channels, many of which exhibit counterparts in X-rays. Throughout this study, \ion{Mg}{ii} h observations are employed for transient detection and localisation. Across the seven flare events analysed in this work, a total of 102 pre-flare transients are identified in the SUIT ROI field of view (see Table~\ref{tab:summary_table}).

\begin{table*}
    \centering
    \begin{tabular}[t]{lccccccr}
    \hline
     Case no. & Number of transients & HEL1OS Counterparts & Most transient location & Nature of flare onset & Hot onset phase\\
     \hline
     Case 1&13&03& Flaring location& Two step & Inconclusive \\
     Case 2&17&08& Flaring location & Two step & Inconclusive \\
     Case 3&12&06& Away from flaring location  & Monotonic rise & Yes \\
     Case 4&19&10& Flaring location & Two step & Yes \\
     Case 5&16&09& Flaring location & Two step & Inconclusive \\
     Case 6&13&08& Away from flaring location & Two step & Yes  \\
     Case 7&12&05& Flaring location & Two step & Yes \\
    \hline
    \end{tabular}
    \caption{Summary of the seven analysed cases. Column (1) lists the case identifier, and column (2) gives the number of pre-flare transients detected in the SUIT \ion{Mg}{ii} h observations. Column (3) shows the number of these transients with X-ray counterparts. Column (4) indicates the location of the dominant transient activity (i.e. the hotspot of transients) relative to the flaring region. 
    Column~(5) describes the morphology of the soft X-ray rise during the flare onset phase. Column~(6) indicates whether a hot-onset precursor 
    event (HOPE) was identified in each case, based on 
    background-subtracted SoLEXS and STIX spectral fitting. 
    Cases 1 and 2 lack suitable STIX observations, and the SoLEXS signal above the background is insufficient for reliable spectral fitting. In Case 5, the background-subtracted signals from both SoLEXS and STIX are too weak for reliable isothermal fitting. Consequently, the hot-onset condition cannot be assessed for these three cases.}
    \label{tab:summary_table}
\end{table*}
\subsection{Spatial distribution of pre-flare transients}

We find that the transients are predominantly concentrated near PILs, consistent with previous studies reporting pre-flare brightenings co-spatial with PILs in AIA 1600 {\AA}, \ion{Ca}{II} H, \ion{Ca}{II} K, X-rays, and H$\alpha$ observations (e.g. \citealt{dissauer_uniqueness_2025, Kusano2012, Bamba2013, Chifor2007, shohinUltravioletXrayPrecursors2024, kumar_compact_2026,wangHighresolutionObservationsFlare2017b}). Evidence from earlier work has also uncovered pre-flare signatures (manifest as H$\alpha$ brightening) in the chromosphere, which tend to overlie regions with a high magnetic twist (chirality) gradient and are close to PIL \citep{Hahn2005}. Together, these results suggest that pre-flare transients are present in the upper chromosphere and are consistent with similar signatures observed at other wavelengths at higher atmospheric layers \citep{ChiMT_2006, Chifor2007, Chitta2026}. However, we do not detect comparable brightening in the continuum filters (2767 {\AA}, 2832 {\AA}, 3000{\AA}, and 3880{\AA}), which suggests that these transient effects do not reach deeper into the photosphere.

Instead of considering one pre-flare brightening just before the flare, we systematically study all pre-flare transients happening in the active region 2 hours before the flare \citep[similar to][]{dissauer_uniqueness_2025} in multiple cases.
We observe that pre-flare transients are not only present along strong PIL, but also present in the plage region and the footpoints of some loops. 
\citet{dissauer_uniqueness_2025} also detected transients in the quiet period, but they are spread across the active region, suggesting that small-scale transients are commonly present supporting \citet{shimizuTransientBrighteningsActive1992a}'s finding. Whereas, during the pre-flare phase, transients seem to be concentrated near the PIL, observed in both this work and \citet{dissauer_uniqueness_2025}.
Based on their spatial association with the flaring region and the PIL in many cases, it is hypothesised that these small-scale transient events may be related to emerging flux or flux cancellation processes \citep{dissauer_uniqueness_2025}. They may progressively destabilise the magnetic arcade, thereby contributing to the conditions necessary for flare onset.
The origin of pre-flare transients and their association with flares requires further detailed study.

\subsection{X-ray counterparts of pre-flare transients}
Of the 102 detected UV transients, 49 events (47\%\ ) exhibit X-ray counterparts in HEL1OS. Using AIA 131~{\AA} observations, we verify the source locations of these X-ray enhancements and find that only 29 events (28\%\ ) originate from the same active region as the SUIT ROI. The remaining 20 X-ray transients are co-temporal but spatially offset, indicating contributions from neighbouring active regions. In some cases, weak brightenings are also observed in the target region, possibly reflecting magnetic connectivity between active regions. The remaining 72\%\ of UV transients do not show associated X-ray counterparts. This may indicate that these events are either too weak to produce detectable X-ray emission or that reconnection occurs at lower atmospheric heights.

STIX spectral data are available for five of the seven flare cases, within which ten pre-flare transient events are identified as ``flare-like'', based on the presence of statistically significant non-thermal emission. This confirms that some of these transients are coronal reconnection events and are strong enough to produce non-thermal emission. \citet{wangHighresolutionObservationsFlare2017b} and \citet{kumar_compact_2026} interpret pre-flare transients as signature of  chromospheric reconnection. Based on our observations, it is difficult to confirm whether the remaining transients are reconnections and, if so, at what height in the solar atmosphere they occur. 

The temperature calculated from SoLEXS shows tiny variations, but they are within the error bars. This could be due to SoLEXS observing the sun{--}as{--}a{--}star and probing the thermal plasma all over the disc. However, \cite{shohinUltravioletXrayPrecursors2024} reports temperature enhancements for such transients; this may also be related to the event's strength. 
\citet{dissauer_uniqueness_2025} measured the temperature using Differential Emission Measure (DEM) for the ROI itself, but the temperature and emission measure enhancement region were not co-spatial.
Although the temperature measured by STIX shows a strong enhancement (above 10 MK), this difference could be due to the fact that the two instruments probe different temperature plasmas.
SoLEXS measures the emission-measure-weighted temperature of the bulk coronal plasma, while STIX is sensitive exclusively to plasma above $\sim$8~MK \citep{battagliaSTIXXrayMicroflare2021}; the two instruments probe different parts of the differential emission measure distribution \citep{warmuthConstraintsEnergyRelease2016}.

\subsection{Flare onset phase}
In all cases where STIX data were available, low-energy X-ray (4{--}8 keV) emission enhancement starts on average, approximately 3 minutes before the impulsive phase, consistent with previously reported precursor behaviour \citep{veronigRelativeTimingSolar2002}. 
We estimate the temperature and emission measure of the onset phase using both SoLEXS and STIX data.
For Cases 1 and 2, although SoLEXS spectra were available, the signal above the background was insufficient to reliably examine the hot-onset condition. 
Similarly, in Case 5, both STIX and SoLEXS spectra lacked sufficient signal above background for reliable spectral fitting. 
Consequently, the hot-onset condition could not be confirmed in these cases.
In the remaining cases, we observe elevated and nearly constant temperatures above 10~MK, together with a monotonic increase in emission measure during the thermal-emission phase, consistent with the HOPE behaviour, as previously described by \citet{hudson_hot_2021} and \citet{battaglia_existence_2023}.

In all cases, the temperatures and emission measures derived from STIX are systematically higher than those obtained from SoLEXS. This difference is likely attributable to the different energy ranges used for spectral fitting, causing the two instruments to sample different portions of the differential emission measure (DEM) distribution, as discussed in the previous section.
Except for the sigmoid flare (Case 3), all cases studied in this paper show a two-step flare initiation: a slow rise in soft X-ray followed by a plateau, and then an impulsive phase, similar to the event described in \citet{awasthiChromosphericResponsePrecursor2018}. 
These authors interpret this as reconnection-triggered non-thermal electron heating in the first step, where non-thermal particles are thermalised in the upper chromosphere or produce emission too faint to detect, followed by a second, impulsive phase in which the main flare occurs. Our X-ray spectral fitting shows that the emission is initially thermal, gradually developing a non-thermal component, supporting this two-step interpretation.
The thermal-to-non-thermal transition is smooth, suggesting that the initially thermal character of the early phase may reflect a sensitivity limitation, or that reconnection-accelerated particles are thermalised in the upper chromosphere, as hypothesised by \citet{awasthiChromosphericResponsePrecursor2018}. Alternative coronal heating mechanisms, such as Alfvén wave dissipation or low-atmosphere turbulence, have also been proposed in this context \citep{fletcherImpulsivePhaseFlare2008a,jeffreyDevelopmentLoweratmosphereTurbulence2018}.
As soft X-ray emission rises, \ion{Mg}{ii}~h also begins to brighten, with a rapid enhancement following the onset of the impulsive phase. No systematic timing offset is observed between the hot onset or the non-thermal emission start time and the \ion{Mg}{ii}~h response; the delay between X-ray and \ion{Mg}{ii}~h is at most 90 seconds, consistent with the cadence uncertainty. Similar pre-flare brightenings at flare footpoints have also been reported in AIA 1600~{\AA} observations \citep[e.g.,][]{battaglia_existence_2023}.
In our study, we are unable to distinguish the signatures to rule out either mechanism.
\subsection{Energies of pre-flare transients and their statistical distribution}
The CCDF shown in Fig.~\ref{fig:ccdf_mle} exhibits three distinct segments. The transients in the first segment correspond to low-intensity events, most of which lack X-ray counterparts. These are likely weak responses to coronal loop activity or to transient events occurring in neighbouring active regions, and are therefore excluded from further interpretation as flare-like events.
In contrast, the transients in the second and third segments exhibit stronger UV emission and resemble small flare-like events. Under the isotropic emission, the estimated luminosities for these events range from $8.7 \times 10^{26} erg~s^{-1}$ to  $8.04 \times 10^{27} erg~s^{-1}$. Owing to SUIT's limited cadence, the full temporal evolution of these events is not well sampled; therefore, these values are expected to underestimate the true peak luminosity.
Assuming a characteristic timescale of 1 minute, the corresponding radiated energies in \ion{Mg}{ii}~h from chromosphere are of the order of $10^{28} erg$ to  $10^{29} erg$. 
This range is similar to the energy scale of micro-flares observed in soft X-rays from the corona ($10^{26} erg$ to  $10^{29} erg$; \citealt{shimizuEnergeticsOccurrenceRate1995a, kotaniThermodynamicPropertiesSmall2023a}). Since chromospheric emission is a fraction of the total radiated energy across different layers \citep{shimizuSimultaneousALMAHinode2021}, these events are likely comparable to typical A- or B-class flares, which can not be detected during this epoch of solar maximum because of the elevated GOES soft X-ray background. 

The statistical distribution of the pre-flare transients follows a broken power-law behaviour, with indices of approximately 1.64 and 3.12 below and above the break, respectively. The higher-energy slope is consistent with the Ly$\alpha$ flare distribution reported by \cite{milligan_lyman-alpha_2020}, the lower-energy slope is comparable to X-ray flare distributions reported in previous studies (e.g. \citealt{masonCoronalHeatingDetermined2023}, \citealt{shimizuEnergeticsOccurrenceRate1995a}), although differences between chromospheric and coronal diagnostics are expected due to variations in emission processes and formation heights.
The broken power-law distribution, with its steep upper slope ($\alpha_2$ = 3.12) and characteristic break energy, indicates that these pre-flare transients do not follow the single power-law behaviour characteristic of solar flares observed in soft X-ray emission \citep{aschwanden25YearsSelfOrganized2016, shimizuEnergeticsOccurrenceRate1995a}, suggesting a physical distinction between pre-flare chromospheric transients and the general coronal flare population. The origin of the first power-law index and break energies is unclear. 

We note that the present analysis is based on a limited sample size, and therefore, the inferred statistical properties should be interpreted with caution. A larger sample of events will be required to establish the robustness of the derived power-law behaviour.
\section{Conclusion}
\label{sec:conclusions}
This study presents the first systematic characterisation of chromospheric pre-flare transients using spatially resolved \ion{Mg}{ii}~h observations from SUIT onboard Aditya-L1, combined with simultaneous X-ray diagnostics from HEL1OS, SoLEXS, and STIX — a multi-wavelength capability unavailable to previous pre-flare studies.

The results presented above provide a comprehensive view of chromospheric and coronal activity during the pre-flare phase. The observed spatial distribution, X-ray associations, and statistical properties of the transients offer new insights into the processes that precede major solar flares. The main conclusions of this work are summarised below.
\begin{enumerate}
\item Frequent small-scale transients are observed in the chromosphere during the pre-flare phase, indicating enhanced low-atmosphere activity prior to flare onset.
\item X-ray counterparts for a subset of events demonstrate coupled chromosphere–corona energy release before major flares.
\item  The strong spatial association of transients with the polarity inversion line suggests a magnetic reconnection-driven origin and a possible role in flare triggering.
\item Power-law behaviour of transient intensities comparable with Lyman alpha flare power law index of 2.82$\pm$ 0.27 reported by \citep{milligan_lyman-alpha_2020}, consistent with A-, B-  or micro-flare statistics.
\end{enumerate}
 
These results improve our understanding of pre-flare transients and their association with solar flares, providing useful inputs for models of flare initiation. The unique capability of SUIT onboard Aditya-L1 to provide spatially resolved NUV chromospheric imaging simultaneously with X-ray diagnostics from HEL1OS and SoLEXS opens a new window for systematic investigations of pre-flare activity. Future work combining these observations with complementary instruments will help establish the physical mechanisms driving pre-flare transients and their role in flare initiation.

\section*{Acknowledgements}
 We thank the referee for helpful comments that improved this paper.
 AHN acknowledges the Dr T.M.A Pai Fellowship offered for the PhD program from Manipal Academy of Higher Education (MAHE), Manipal, and the SPARC III program of MoE, Govt of India, coordinated by IIT Kharagpur, for supporting advanced training at the Centre for Astrophysics $|$ Harvard \& Smithsonian, USA. AHN also thanks members of the Solar and Stellar X-ray Group (SSXG) at the Centre for Astrophysics $|$ Harvard \& Smithsonian, USA, for valuable discussions and insightful input during this work. SP and AHN acknowledge the Manipal Centre for Natural Sciences, MAHE, for its facilities and support. SR is supported by the ANRF NPDF Grant No. PDF/2025/005797.
 This work is partly supported by ISRO/RESPOND project “Solar Flares: Physics and Forecasting for Better Understanding of Space Weather”, ISRO/RES/2/438/22-23.

Aditya-L1 is an observatory class mission which is fully funded and operated by the Indian Space Research Organisation (ISRO). The mission was conceived and realised with the help from various ISRO centres. The science payloads and science ready data products are realised by the payload PI institutes in close collaboration with ISRO centres. The PI institutes are: Indian Institute of Astrophysics (IIA); Inter University Centre for Astronomy and Astrophysics (IUCAA), Laboratory for Electro-optics Systems (LEOS/URSC); Physical Research Laboratory (PRL); U R Rao Satellite Centre (URSC); and Space Physics Laboratory (SPL/VSSC). We acknowledge the use of data from the Aditya-L1 mission of the ISRO, archived at the Indian Space Science Data Centre (ISSDC).
SUIT is built by a consortium led by the IUCAA, Pune, and supported by ISRO as part of the Aditya-L1 mission. The consortium comprises CESSI-IISER Kolkata (Ministry of Education), IIA, MAHE, MPS, USO/PRL, and Tezpur University. 
SoLEXS and HEL1OS are designed and developed at the Space Astronomy Group of U R Rao Satellite Centre (URSC), Indian Space Research Organisation, with the help of various entities within URSC. The Space Astronomy Group of URSC also operates the Payload Operation Centre where the higher level data product is generated and posted to ISSDC for dissemination to the scientific community.
SDO data are courtesy of the NASA/SDO AIA and HMI science teams.
The STIX instrument is an international collaboration between Switzerland, Poland, France, Czech Republic, Germany, Austria, Ireland, and Italy.

\section*{Data Availability}
All Aditya-L1 data used in this study are publicly available on the ISSDC-Pradan website \href{https://pradan1.issdc.gov.in/al1/}{https://pradan1.issdc.gov.in/al1/}.
SDO/AIA and SDO/HMI data are available in \href{http://jsoc.stanford.edu/ajax/lookdata.html}{JSOC, Stanford}. 
The STIX data used in this research are available from the \href{https://datacenter.stix.i4ds.net/}{STIX data centre}.



\bibliographystyle{mnras}
\bibliography{citations} 


\appendix
\section{Multi-wavelength light curves for Cases 1, 2, 5, 6, and 7}
Multi-wavelength light curves for the five cases not presented as individual case studies in Section~\ref{sec:results} are shown here for completeness. 
Each figure follows the same format as Fig.~\ref{fig:c7_lc}. 
The pre-flare transients identified in each case are summarised 
in Table~\ref{tab:summary_table}.
\begin{figure}
    \centering
    \includegraphics[width=1\linewidth]{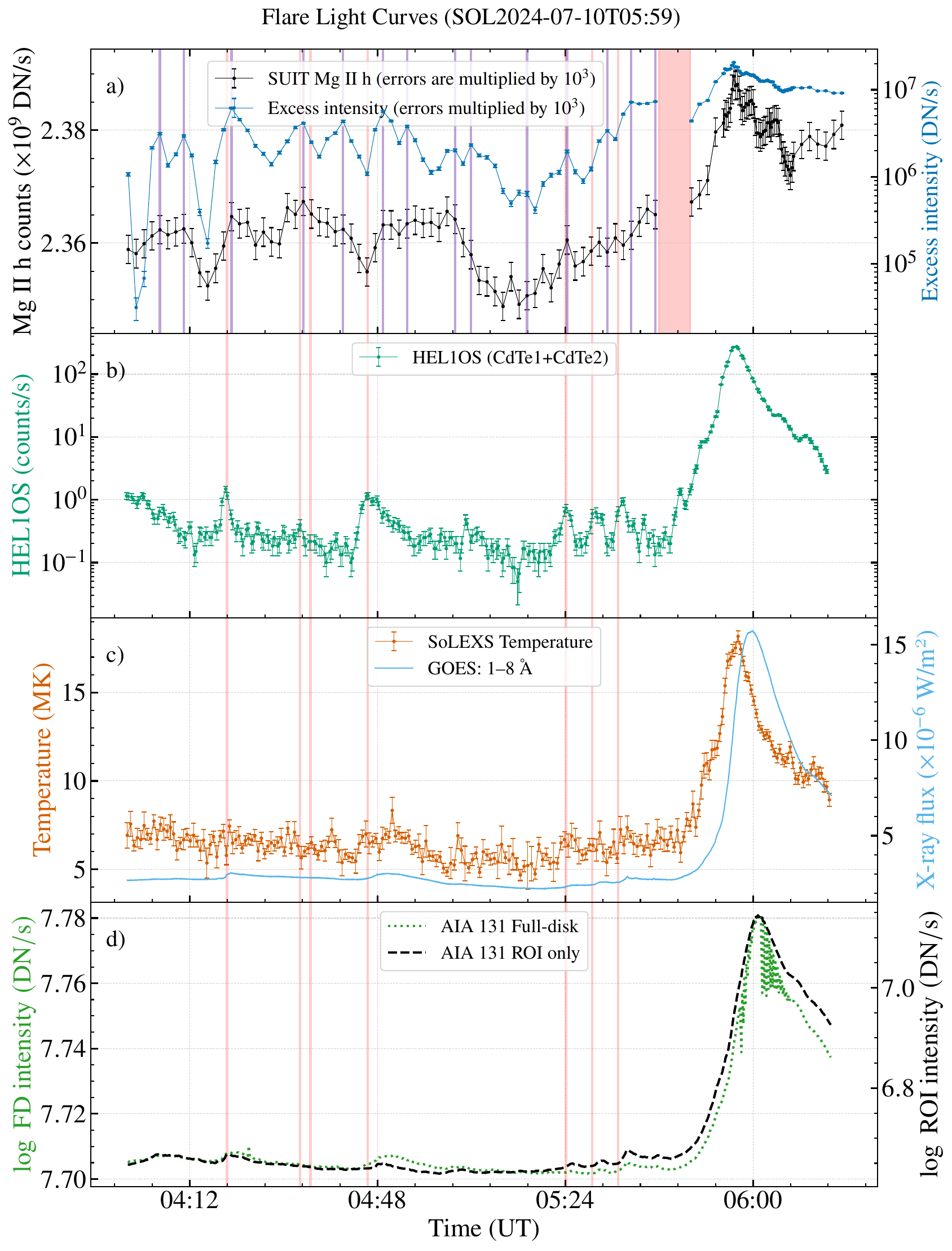}
    \caption{Case 1 Light curves; a) Total intensity light curve of the \ion{Mg}{ii} h ROI images (black). The errors of the individual data points are scaled by $10^3$ for visibility. The light curve of the difference-image intensity above the threshold is shown in cyan, with errors scaled by $10^3$ for clarity. b) HEL1OS light curve (CdTe 1 and CdTe 2, 10{--}30 keV), binned to 1-minute intervals, c) Temperature derived from SoLEXS(orange) and X-ray flux from the GOES 1{--} 8 {\AA} band (sky blue), (d) Total intensity light curve from AIA 131 {\AA}~ integrated up to 1.1 \(R_\odot\) (green dot line), together with the AIA 131~{\AA} light curve extracted from the region corresponding to the SUIT ROI (black dashed line). The magenta vertical lines represent transients identified in \ion{Mg}{ii} h, while the red vertical lines represent transients identified from HEL1OS. The red shaded band indicates time intervals during which SUIT observations are unavailable.}
    \label{fig:c4_lc}
\end{figure}

\begin{figure}
    \centering
    \includegraphics[width=1\linewidth]{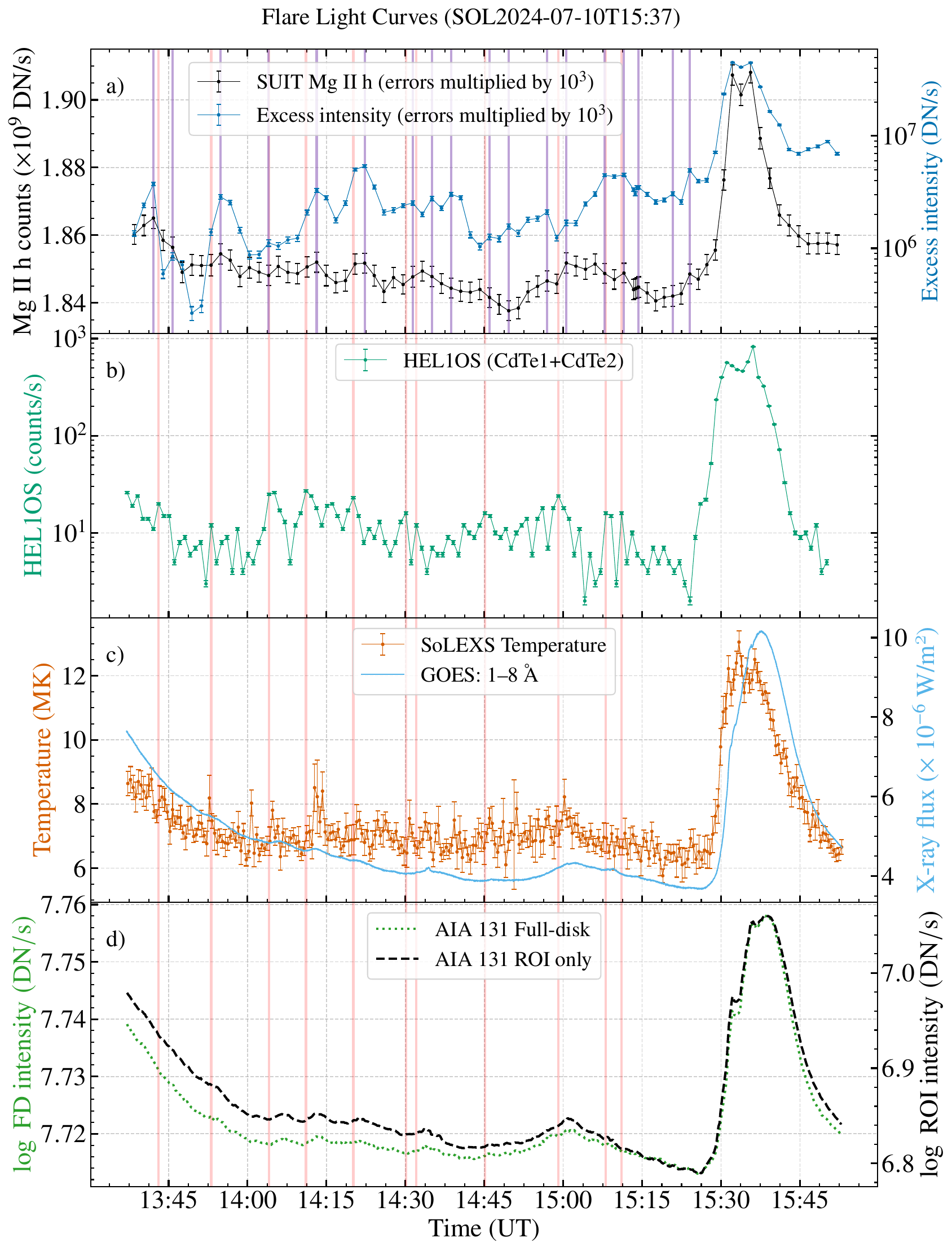}
    \caption{Case 2 Light curves; a) Total intensity light curve of the \ion{Mg}{ii} h ROI images (black). The errors of the individual data points are scaled by $10^3$ for visibility. The light curve of the difference-image intensity above the threshold is shown in cyan, with errors scaled by $10^3$ for clarity. b) HEL1OS light curve (CdTe 1 and CdTe 2, 10{--}30 keV), binned to 1-minute intervals, c) Temperature derived from SoLEXS(orange) and X-ray flux from the GOES 1{--} 8 {\AA} band (sky blue), (d) Total intensity light curve from AIA 131 {\AA}~ integrated up to 1.1 \(R_\odot\) (green dot line), together with the AIA 131~{\AA} light curve extracted from the region corresponding to the SUIT ROI (black dashed line). The magenta vertical lines represent transients identified in \ion{Mg}{ii} h, while the red vertical lines represent transients identified from HEL1OS. The red shaded band indicates time intervals during which SUIT observations are unavailable.}
\end{figure}
\begin{figure}
    \centering
    \includegraphics[width=1\linewidth]{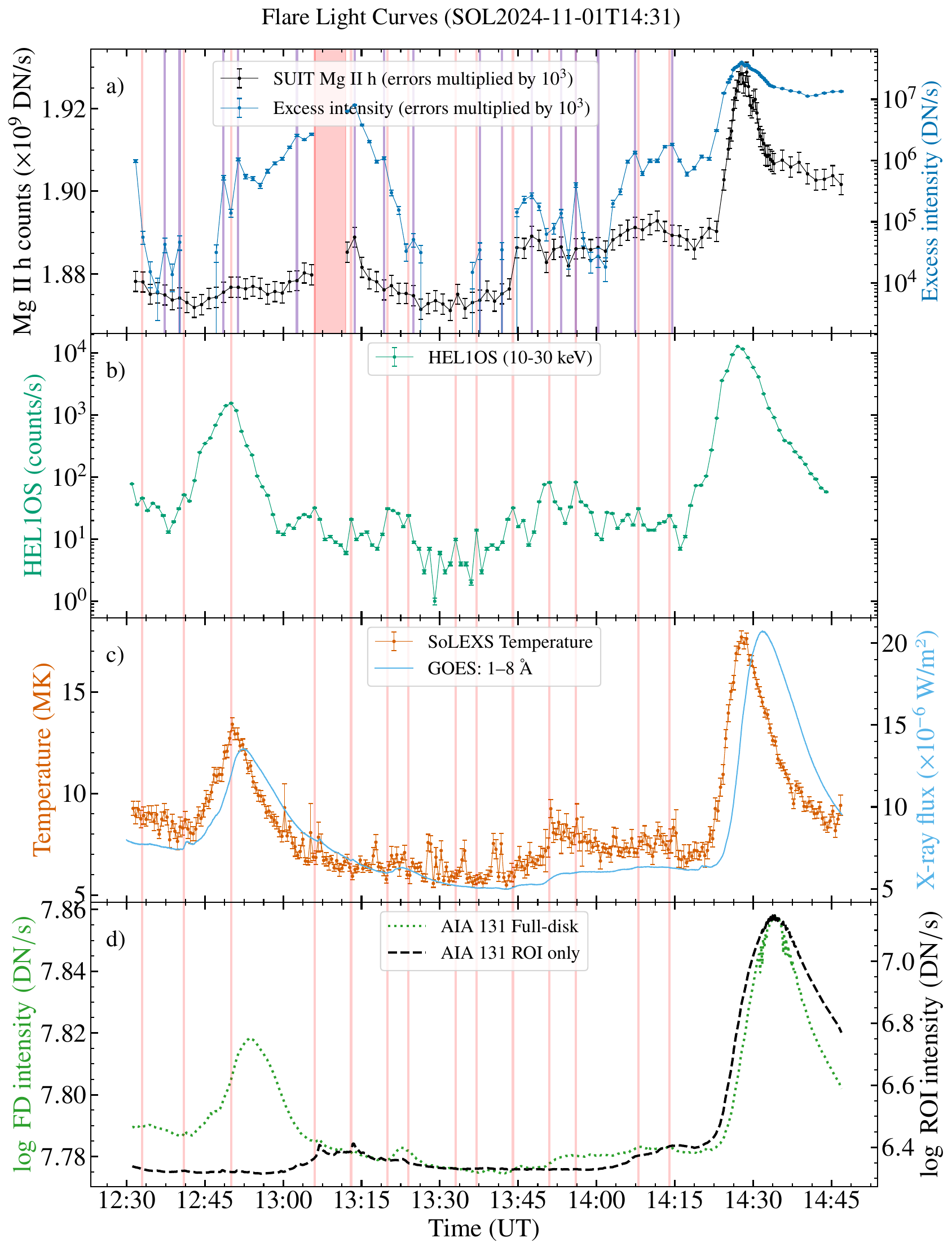}
    \caption{Case 5 Light curves; a) Total intensity light curve of the \ion{Mg}{ii} h ROI images (black). The errors of the individual data points are scaled by $10^3$ for visibility. The light curve of the difference-image intensity above the threshold is shown in cyan, with errors scaled by $10^3$ for clarity. b) HEL1OS light curve (CdTe 1 and CdTe 2, 10{--}30 keV), binned to 1-minute intervals, c) Temperature derived from SoLEXS(orange) and X-ray flux from the GOES 1{--} 8 {\AA} band (sky blue), (d) Total intensity light curve from AIA 131 {\AA}~ integrated up to 1.1 \(R_\odot\) (green dot line), together with the AIA 131~{\AA} light curve extracted from the region corresponding to the SUIT ROI (black dashed line). The magenta vertical lines represent transients identified in \ion{Mg}{ii} h, while the red vertical lines represent transients identified from HEL1OS. The red shaded band indicates time intervals during which SUIT observations are unavailable.}
    \label{fig:c8_lc}
\end{figure}
\begin{figure}
    \centering
    \includegraphics[width=1\linewidth]{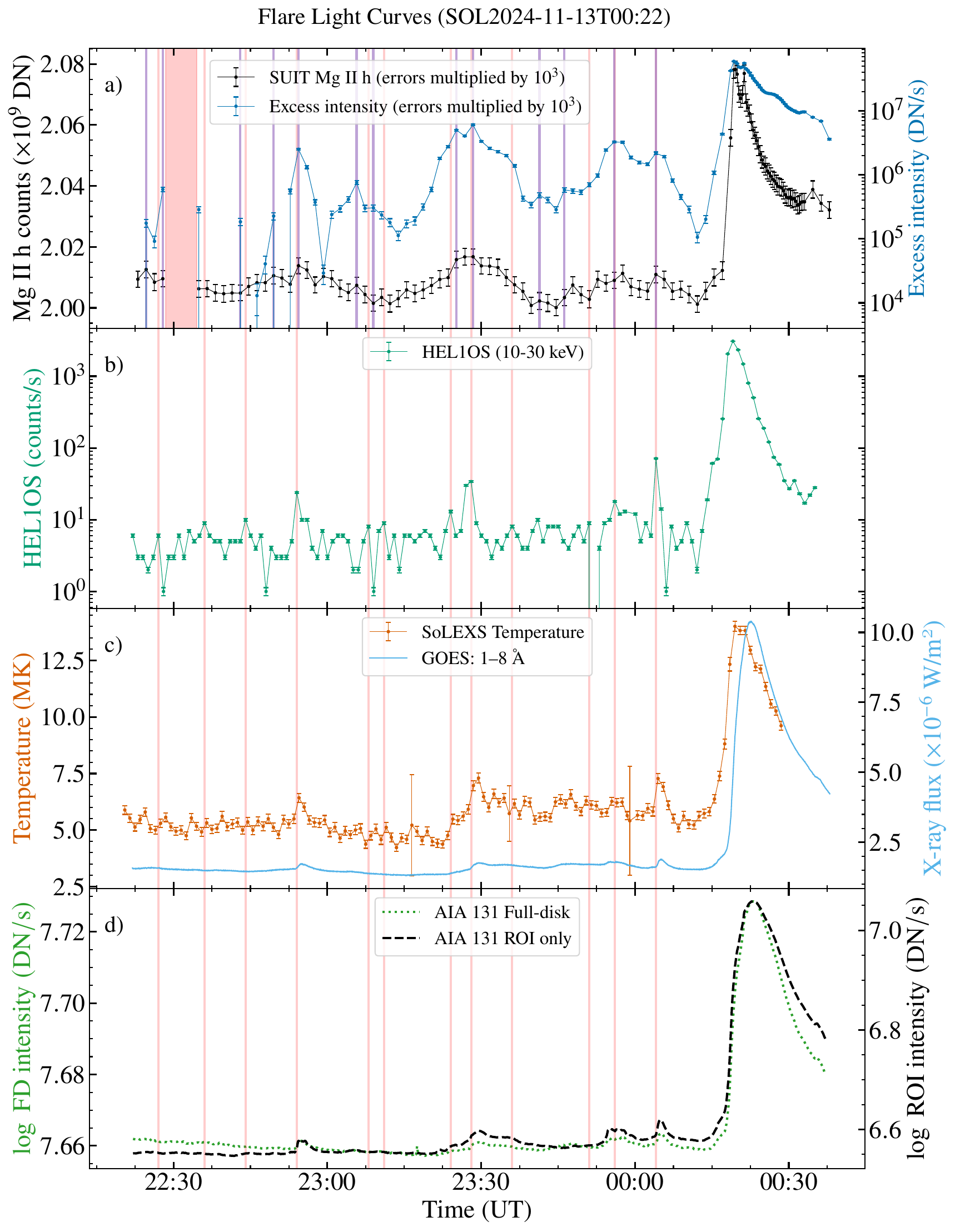}
    \caption{Case 6 Light curves; a) Total intensity light curve of the \ion{Mg}{ii} h ROI images (black). The errors of the individual data points are scaled by $10^3$ for visibility. The light curve of the difference-image intensity above the threshold is shown in cyan, with errors scaled by $10^3$ for clarity. b) HEL1OS light curve (CdTe 1 and CdTe 2, 10{--}30 keV), binned to 1-minute intervals, c) Temperature derived from SoLEXS(orange) and X-ray flux from the GOES 1{--} 8 {\AA} band (sky blue), (d) Total intensity light curve from AIA 131 {\AA}~ integrated up to 1.1 \(R_\odot\) (green dot line), together with the AIA 131~{\AA} light curve extracted from the region corresponding to the SUIT ROI (black dashed line). The magenta vertical lines represent transients identified in \ion{Mg}{ii} h, while the red vertical lines represent transients identified from HEL1OS. The red shaded band indicates time intervals during which SUIT observations are unavailable.}
    \label{fig:c9_lc}
\end{figure}
\begin{figure}
    \centering
    \includegraphics[width=1\linewidth]{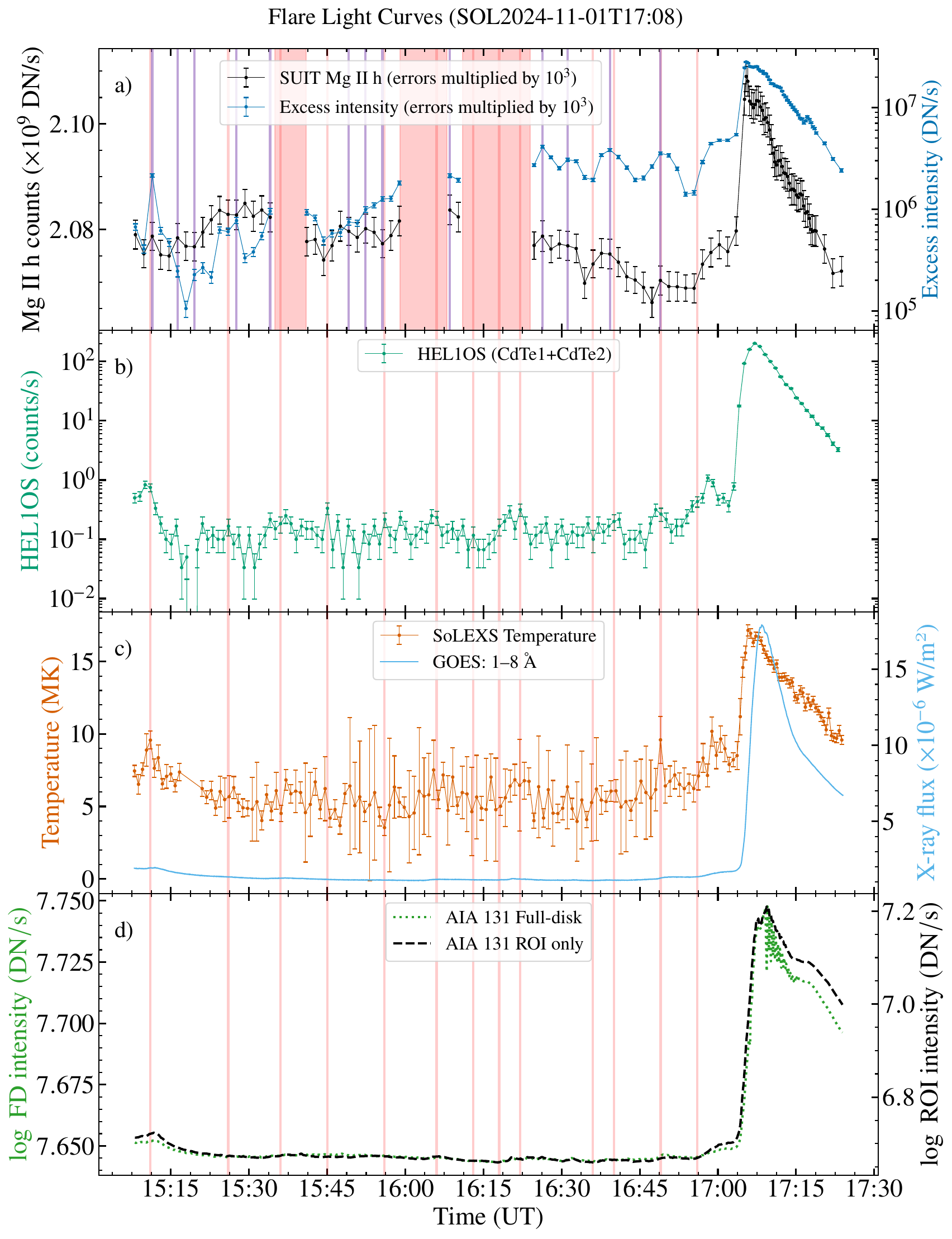}
    \caption{Case 7 Light curves; a) Total intensity light curve of the \ion{Mg}{ii} h ROI images (black). The errors of the individual data points are scaled by $10^3$ for visibility. The light curve of the difference-image intensity above the threshold is shown in cyan, with errors scaled by $10^3$ for clarity. b) HEL1OS light curve (CdTe 1 and CdTe 2, 10{--}30 keV), binned to 1-minute intervals, c) Temperature derived from SoLEXS(orange) and X-ray flux from the GOES 1{--} 8 {\AA} band (sky blue), (d) Total intensity light curve from AIA 131 {\AA}~ integrated up to 1.1 \(R_\odot\) (green dot line), together with the AIA 131~{\AA} light curve extracted from the region corresponding to the SUIT ROI (black dashed line). The magenta vertical lines represent transients identified in \ion{Mg}{ii} h, while the red vertical lines represent transients identified from HEL1OS. The red shaded band indicates time intervals during which SUIT observations are unavailable.}
    \label{fig:c10_lc}
\end{figure}

\section{AIA 131~\AA\ regional light curves for X-ray source localisation}

\begin{figure*}
    \centering
    \includegraphics[width=1\linewidth]{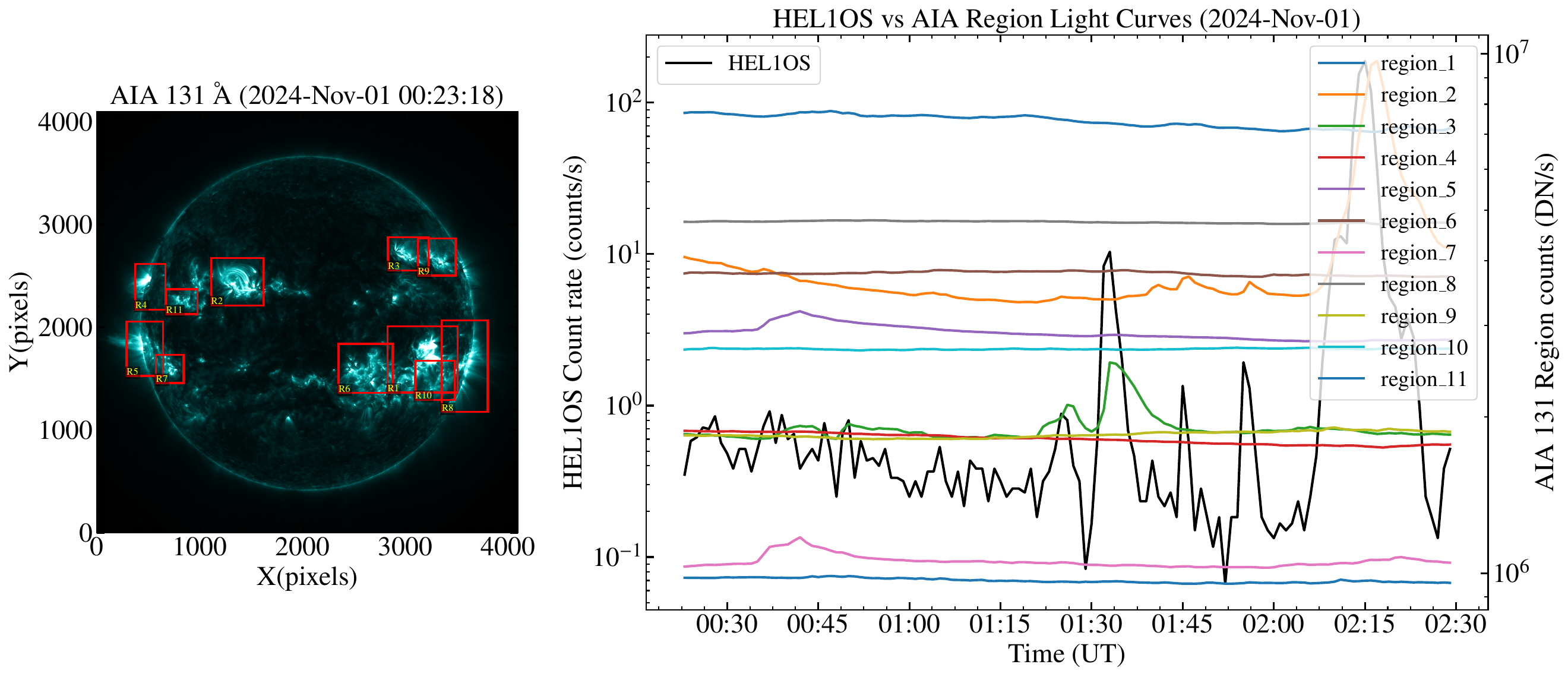}
    \caption{AIA 131 {\AA} image of the 2024 November 1 event (left), with active region boundaries marked. Right: AIA 131 {\AA} light curves for each identified region (coloured lines) compared with the HEL1OS CdTe 10--30~keV count rate (black line)}
    \label{fig:aia_131_all_lc}
\end{figure*}

Since SoLEXS and HEL1OS observe the Sun as a star, they cannot spatially resolve the source of individual X-ray enhancements. To determine whether a given 
enhancement originates from the target active region 
observed by SUIT, or from another active region 
elsewhere on the solar disk, we extract AIA 131~\AA\ 
light curves for all active regions visible during 
entire pre-flare phase. 
Each light curve represents the total AIA 131~\AA\ intensity integrated over a selected region enclosing the relevant active region. 
Fig.~\ref{fig:aia_131_all_lc} shows an example for the 2024 November 1 observations (case 4): the coloured lines show the AIA 131 {\AA} light curves for each identified 
region, compared with the HEL1OS CdTe (10--30~keV) 
light curve (black). A given X-ray enhancement is attributed to the SUIT target region when a clear, corresponding brightening is visible primarily in the target active-region light curve and is absent or substantially weaker in the light curves of all other identified regions


\bsp	
\label{lastpage}

\end{document}